\documentclass[epj,twocolumn=false]{svjour} 
\usepackage{graphics}
\usepackage[colorlinks=true,linkcolor=blue, citecolor=blue, urlcolor=blue]{hyperref}
\usepackage{color}
\usepackage{multirow}
\usepackage{amssymb}
\usepackage{amsmath}
\usepackage{slashed}
\usepackage{graphicx}
\usepackage{placeins}
\usepackage{array}
\usepackage{dblfloatfix}
\usepackage{scalerel} 
\usepackage{xcolor}
\usepackage{float}
\usepackage{multirow}
\usepackage{stackengine}
\usepackage{siunitx}

\usepackage{dcolumn}
\usepackage{bm}
\usepackage{xspace}
\usepackage{hyperref}

\newcommand{\ptjet}{\mbox{$p_{{\mathrm{T}}}^{\mathrm{jet}}$}}
\newcommand{\mr}[1]{\mathrm{#1}}
\newcommand{\pt}{\mbox{$p_{{\mathrm{T}}}$}}
\newcommand{\cf}{c_{\mathrm{F}}}
\newcommand{\Kt}{\mbox{$k_{t}$}}
\newcommand{\mathptjet}{p_{\mathrm{T}}^{\mathrm{jet}}}
\newcommand{\mathptz}{{p_\mathrm{T}}_0}
\newcommand{\ptz}{\mbox{${p_\mathrm{T}}_0$}}
\newcommand{\ptzprime}{\mbox{${p'_\mathrm{T}}_0$}}
\newcommand{\fiz}{\mbox{$f_{i,0}$}}
\newcommand{\Raa}{\ensuremath{R_\mathrm{AA}}\xspace}
\newcommand{\pars}{\ensuremath{\langle \Delta p_\mathrm{T}^\mathrm{jet} \rangle} \xspace}
\newcommand{\parl}{\ensuremath{\langle L \rangle} \xspace}
\newcommand{\npart}{\ensuremath{N_{\mathrm{part}}} \xspace}

\onecolumn
\begin{document}

\title{Flavor and path-length dependence of jet quenching from inclusive jet and $\gamma$-jet suppression}

\author{Agnieszka Ogrodnik 
 \and
Martin Rybář 
 \and
Martin Spousta 
\mail{martin.spousta@matfyz.cuni.cz}
}

\institute{
 Institute of Nuclear Physics, Faculty of Mathematics and Physics, Charles University 
 }

\date{Received: date / Revised version: date}

\abstract{
We employ the parametric approach to analyze jet suppression measured using the nuclear modification factor of inclusive jets, $b$-jets, and jets from $\gamma$-jet events. With minimum model assumptions, we quantify the magnitude of the average energy loss and its transverse momentum dependence. Then, we quantify the impact of fluctuations in the energy loss and nuclear PDFs on the measured jet suppression. The Glauber and Trajectum models are used to estimate the average path length that a jet is expected to traverse in the medium. With this information, we quantify the path-length dependence of the average energy loss, which is found to support the physics picture of the radiative nature of parton energy loss. 
Using the obtained parameterizations, we evaluate jet $v_2$, quantifying its sensitivity to details of path-length modeling.
We also provide model-independent predictions for the magnitude of energy loss expected in upcoming oxygen-oxygen collisions and briefly discuss the path-length dependence of energy loss of $b$-jets. Finally, we analyze intriguing features seen in the suppression of jets in the $\gamma$-jet system. 
}

\authorrunning{A. Ogrodnik et al.}
\titlerunning{Flavor and path-length dependence of jet quenching from inclusive jet and $\gamma$-jet suppression}

\maketitle

\newpage
\tableofcontents

\section{\label{sec:intro} Introduction}

Calculations of lattice quantum chromodynamics (QCD) predict the emergence of deconfined matter in ultra-relativistic collisions of heavy ions \cite{Ding:2015ona}. This type of matter, composed of quarks and gluons freed from hadrons, which is also called quark-gluon plasma (QGP), existed during the first microseconds of the evolution of the universe \cite{Busza:2018rrf}. It is, therefore, of importance to study the properties of QGP and the mechanisms of its interaction with elementary particles passing through it. Such studies may be performed using jets of hadrons originating in a hard scattering of elementary quarks and gluons occurring within the volume of QGP medium. Jets are described in perturbative QCD as virtuality-ordered and/or angular-ordered showers of partons propagating through the vacuum.
  When passing the QGP medium, partons from the parton shower are assumed to lose energy predominantly by medium-induced radiative processes \cite{Mehtar-Tani:2013pia,Majumder:2010qh,Casalderrey-Solana:2007knd,Cao:2020wlm}. This
leads to modification of yields and structure of jets with respect to
those passing only the vacuum \cite{Apolinario:2022vzg,Cunqueiro:2021wls,Connors:2017ptx}. This phenomenon is commonly termed jet quenching.

One of the important open questions related to jet quenching is how much the energy loss depends on the color charge of the initial parton \cite{Citron:2018lsq,Cunqueiro:2021wls}. This color charge dependence may have two qualitatively different origins. 
First, it may originate as a consequence of the structure of the parton shower, which is, on average, different for quark-initiated and gluon-initiated jets \cite{Dissertori2009,Gallicchio:2011xq}. Quark-initiated jets, which typically produce harder and narrower parton showers with fewer constituents, may naturally be expected to lose less energy than gluon-initiated jets, which have softer and wider parton showers with more constituents.
  This principle is encoded in many theoretical descriptions in different ways, compare e.g. Refs.~\cite{Mehtar-Tani:2018zba,Chien:2015hda,Casalderrey-Solana:2015vaa}, and it is also implemented in jet quenching Monte-Carlo generators \cite{Zapp:2008gi,Armesto:2009fj,Casalderrey-Solana:2014bpa,Schenke:2009gb,Majumder:2013re,Wang:2013cia}.
Second, the color charge dependence of the energy loss may be connected with the phenomena of color (de)coherence \cite{Mehtar-Tani:2010ebp,Casalderrey-Solana:2012evi}. In that picture, a substantial part of the parton shower loses the energy coherently. The energy loss is then driven by a color charge of the initial quark or gluon. Besides analytical calculations of color coherent energy loss \cite{Mehtar-Tani:2010ebp,Mehtar-Tani:2011hma,Mehtar-Tani:2011vlz,Casalderrey-Solana:2012evi,Apolinario:2014csa,Mehtar-Tani:2017web,Barata:2021byj}, the color coherence starts to be discussed in the context of Monte-Carlo (MC) simulations of jet quenching as well \cite{Caucal:2019uvr,JETSCAPE:2022jer,Cunqueiro:2023vxl}.

In addition to the incoherent and color-coherent regimes of in-medium emissions, an admixture of vacuum-like emissions \cite{Caucal:2018dla} may also contribute to the overall observable jet quenching. To address the original question on the role of jet flavor in this complex physics setup, detailed studies comparing observables having different sensitivity to the flavor of the initial parton need to be performed. One example of such studies is the study suggesting measurements of forward jet substructure \cite{Pablos:2022mrx}, which, however, have not yet been measured in Pb+Pb collisions at the LHC. Another possibility to study the flavor sensitivity may come from measurements of jets originating in $\gamma$-jet final state \cite{Wang:1996yh}, which offer a sample dominated by quark-initiated jets \cite{ATLAS:2023iad}. For this final state, multiple measurements were performed at the LHC \cite{ATLAS:2023iad,CMS:2012ytf,CMS:2017ehl,CMS:2018jco,CMS:2018mqn,ATLAS:2018dgb,ATLAS:2019dsv,Liu:2022yjc}. Understanding the jet production in the $\gamma$-jet final state provides not only a tool to study the flavor dependence, but it should also allow a better understanding of the role of selection biases and the path-length dependence of the energy loss \cite{Zhang:2009rn,Qin:2009bk,Renk:2006qg}. Here, in particular, the quantitative assessment of path-length dependence of parton energy loss remains an open issue for a long time \cite{Betz:2014cza,Betz:2016ayq,Djordjevic:2018ita,Arleo:2022shs}. 
Studying similarities and differences between the inclusive jet suppression and suppression of jets from $\gamma$-jet final state is therefore of crucial importance.

In this paper, we apply the parametric approach to jet quenching (also dubbed ``EQ-model'') \cite{Spousta:2015fca,Spousta:2016agr} to benchmark the impact of different components contributing to the final observable jet quenching. The paper is organized as follows. Section~\ref{sec:mod} introduces and extends the parametric modeling of jet quenching. Section~\ref{sec:spec} discusses the simulation of input MC samples used in the study. Section~\ref{sec:incl} quantifies the magnitude of the jet quenching in inclusive jets and compares the ability of different parametric implementations to reproduce measured jet $\Raa$. Section~\ref{sec:path} extracts the path-length dependence of jet suppression using Glauber and Trajectum models. Given the path-length dependence, the jet $v_2$ is studied, pointing to its sensitivity to details of path-length modeling. 
Then, predictions for jet $\Raa$ in oxygen-oxygen collisions are provided,  and the 
differences between the inclusive jet suppression and suppression of $b$-jets are briefly discussed. In Section~\ref{sec:gamma}, extracted parameters of jet quenching from inclusive jet $\Raa$ are used to study the suppression of jets in the $\gamma$-jet system. The last section then provides a summary and conclusions.

\section{Parametric approach to parton energy loss}
\label{sec:mod}

In Ref. \cite{Spousta:2015fca}, the parametric approach to the jet quenching modeling was introduced. The basic ingredient of the modeling is the approximation in which observable jet spectra consist of two components: quark-initiated jet spectra and gluon-initiated jet spectra. Individual spectra can be precisely parameterized using the modified power law, and combined spectra can then be expressed as,
\begin{equation}
     \frac{dN(\ptjet)}{d\ptjet} =  A \sum_{i=1}^2 \fiz \left( \frac{\ptz}{\ptjet}\right)^{n_i(\ptjet)},
\label{eq:unmod}
\end{equation}
where $A$ is the normalization factor, $\ptz$ is an arbitrary parameter to keep the exponent dimensionless (here and in what follows, $\ptz = 40$~GeV is used). The index $i=1,2$ labels quark-initiated and gluon-initiated jets, respectively, and $\fiz$ is a fraction of quark-initiated or gluon-initiated jets at $\ptz$ ($f_{1,0}=1-f_{2,0}$). The exponent $n_i$ is \ptjet\ dependent, 
\begin{equation}
  n_i(\ptjet) = \sum_{j=0}^{j_\mathrm{max}} \beta_j \log^j \bigg( \frac{\ptz}{\ptjet} \bigg).
\label{eq:exp}
\end{equation}
Here, $\beta_j$ are free parameters, and $j_\mathrm{max}=3$ was reported to allow sufficient precision in the parameterization of jet spectra.

The average transverse momentum lost by the jet is then expressed as
\begin{equation}
    \pars_i \equiv \pars_i(\ptjet) = c_{F,i} ~ s ~ \bigg( \frac{\ptjet}{\ptz} \bigg)^\alpha,
\label{eq:s}
\end{equation}
where 
$i=1,2$ as in Equation~(\ref{eq:unmod}) and $c_{F,i}$, $\alpha$, and $s$ are free parameters. The color factor $c_{F,1}$ for quark-initiated jets is set to unity, and $c_{F,2}$ for gluon-initiated jets is labeled $c_{F}$.

The quantification of energy loss using Equation~(\ref{eq:s}) then allows to derive the analytical expression for the jet nuclear modification factor,
\begin{equation}
\Raa(\ptjet) =
 \frac{d N_Q(\ptjet)}{d \mathptjet} \bigg/ \frac{d N(\ptjet)}{d\ptjet} = 
 \sum_{i=1}^{2} f_i(\ptjet,n_i,f_{i,0}) ~ g_i(\ptjet,n_i,\pars_i)
\label{eq:raa}
\end{equation}
where subscript $Q$ denotes quenched jet spectra, $g_i$ is an analytical function calculated in~\cite{Spousta:2015fca}, and $f_i$ are the flavor fractions given by
\begin{equation}
f_1 (\ptjet,n_i,f_{i,0}) =
 \left[1 + \frac{f_{2,0}}{f_{1,0}}\left(
 \frac{\mathptjet}{\mathptz}\right)^{n_2(\mathptjet) - n_1(\mathptjet)} \right]^{-1}, ~ f_2 = 1-f_1,
\end{equation}
with $f_{i,0} = f_i(\ptz)$.
The analytical expression for the parameterized \Raa\ then allows for convenient fitting of the data.

In the first study using the LHC run-1 data at $\sqrt{s_\mathrm{NN}} = 2.76$~TeV \cite{Spousta:2015fca}, $c_F$ was fixed to the leading order value of 9/4, and $\alpha=0.55$ was found to describe the data well. In the subsequent study which used rapidity dependent LHC run-1 data \cite{Spousta:2016agr}, following values of parameters were determined: $c_F = 1.78 \pm 0.12$, $\alpha = 0.52 \pm 0.02$, and $s$ depending linearly on $N_\mathrm{part}$.

Besides the analytical approach, MC simulation can also be performed by modifying the momentum of quark- and gluon-initiated jets delivered by the MC generator using Equation~(\ref{eq:s}). Modified spectra for the two flavors can then be combined and divided by the original spectra with no modification in order to obtain the \Raa. The simulated \Raa\ can be compared with the data to obtain parameters minimizing the difference between the data and the simulated \Raa. Initial unmodified jet spectra can either be delivered directly by the MC generator or the output from the MC generator can be parameterized by Equation~(\ref{eq:unmod}) separately in rapidity intervals, and jets may be obtained by sampling from the parameterized spectra. This latter approach is used in this study. The precise parameterization of jet spectra, which also uses reweighting to the measured data in $pp$ collisions, is provided in Section~\ref{sec:spec}. We should stress that parameters characterizing jet spectra do not count as parameters needed to model the jet quenching, since the parametric quenching model can run directly on MC, albeit with worse precision given by the difference in the modeling of $pp$ reference by MC.  

Two-component (quark/gluon) classification of jet quenching, allowing to derive jet $\Raa$, was used in various publications \cite{Takacs:2021bpv,Adhya:2021kws,Pablos:2022mrx,Mehtar-Tani:2022zwf,Caucal:2020uic,Ke:2020clc,Qiu:2019sfj}, proving the usefulness of this approach. 
While still aiming for the jet quenching description with minimal model assumptions and a minimal number of free parameters, we 
intend to go beyond the approach of average energy loss. To do that, we introduce fluctuations 
by convoluting the single jet spectrum with the energy loss distribution $w(\ptjet,\Delta \ptjet)$,  
   \begin{equation}
    \frac{dN_Q(\ptjet)}{d\ptjet} = \int d \Delta \ptjet \frac{d N(\ptjet + \Delta \ptjet)}{d \ptjet} w(\ptjet + \Delta \ptjet, \Delta \ptjet), 
   \label{eq:fluct1}
   \end{equation}
where $\Delta \ptjet$ is connected with the average energy loss by 
   \begin{equation}
    \pars = \int d \Delta \ptjet ~ \Delta \ptjet ~ w(\ptjet, \Delta \ptjet). 
   \label{eq:fluct2}
   \end{equation}
One may further assume, as e.g. in Refs.~\cite{He:2018gks,Zhang:2023oid}, that energy loss fluctuations depend only on $ \Delta \ptjet / \langle \Delta \ptjet \rangle \equiv x$. To model the energy loss distribution, a generalized integrand of gamma function can be used,
   \begin{equation}
    \tilde{w}(x) = \frac{c_1^{c_0}}{\Gamma(c_0)} x^{c_0-1} e^{-c_1 x},
   \label{eq:fluct3}
   \end{equation}
where $c_0$ and $c_1$ are free parameters and (\ref{eq:fluct3}) is connected with (\ref{eq:fluct2}) by relation $x \tilde{w}(x)/\langle \tilde{w}(x)\rangle = \Delta \ptjet w(\ptjet, \Delta \ptjet)$. For $c_0 = c_1$, the functional form (\ref{eq:fluct3}) coincides with functional form used in previous studies ~\cite{He:2018xjv,He:2018gks,Wu:2023azi} that obtained $c_0 = c_1 \approx 4$. Setting $c_0 \neq c_1$ allows parameterizing fluctuations used in the recent boson-jet study \cite{Brewer:2021hmh}. The quenching probability distributions provided in Fig.~3 of \cite{Brewer:2021hmh} were used as an input to fitting by (\ref{eq:fluct3}), leading to values $c_0 = 1.4 \pm 0.2$ and $c_1 = 0.3 \pm 0.1$ (central value describes the case of the quenching with the medium response). This latter parameterization leads to a somewhat narrower spectrum of fluctuations compared to the case of $c_0 = c_1 \approx 4$. 
Implementing fluctuations, therefore, adds one or two additional free parameters to the modeling, which are, however, not a subject of minimization but are fixed by theory.  

Besides introducing parameterization of fluctuations, one may further generalize the power-law functional form to allow logarithmic $\ptjet$ dependence,
   \begin{equation}
    \pars = c_F ~ s ~ \bigg( \frac{\ptjet}{\ptz} \bigg)^\alpha \log \bigg( \frac{\ptjet}{\ptzprime} \bigg),
   \label{eq:s2}
\end{equation}
where $\ptzprime$ was set to 1 GeV. This parameterization keeps the same number of free parameters as (\ref{eq:s}), and it is used, e.g., in studies performed with the linear Boltzmann transport (LBT) model~\cite{He:2018gks}.

\section{Input parton spectra}
\label{sec:spec}

The inclusive jet production used in this study is simulated using PYTHIA8 \cite{Sjostrand:2014zea,Bierlich:2022pfr} with  A14 tune \cite{TheATLAScollaboration:2014rfk} and a NNPDF 2.3 LO PDF set \cite{Ball:2012cx} which was used as a reference in recent experimental studies of jet production in $pp$ collisions at 5.02~TeV \cite{ATLAS:2023hso}. As an alternative, inclusive jets simulated using HERWIG7 \cite{Bahr:2008pv} are used. The PYTHIA and HERWIG samples are often used together in order to provide a systematic check of the flavor dependence of jet production \cite{ATLAS:2020cli,CMS:2016lmd}, which is known to be simulated differently in these two Monte-Carlo (MC) generators. To reconstruct jets, the anti-\Kt\ algorithm \cite{Cacciari:2008gp} implemented in FastJet package \cite{Cacciari:2011ma} is used with the default value of distance parameter $R=0.4$ ($R=0.2$ jets are used in a brief study of $b$-jets). 
The $\gamma$-jet production is simulated using PYTHIA8 with A14 tune and a NNPDF 2.3 LO PDF set. To match the conditions of analysis in Ref.~\cite{ATLAS:2023iad}, photons are required to be isolated, have $p_{T}^{\gamma}>50$~GeV, and $|\eta|<2.37$. Photons are isolated if the sum of transverse energy within the $R=0.3$ cone around the photon is less than 3 GeV. Selected jets have $\ptjet > 50$~GeV, $|\eta|<2.8$, and are opposite to selected photons in the azimuthal angle ($\Delta\phi_{\gamma,\mathrm{jet}}>7\pi/8$). The $\gamma$-jet cross-section measured in $pp$ collisions has two contributions, one from photons originating directly in the hard scattering (`prompt photons') and one from photons originating in the fragmentation of a jet in a dijet system (`fragmentation photons'). In the simulation by PYTHIA8, the latter contribution constitutes 35\% of the total cross-section for jets within the studied kinematic range and cannot, therefore, be neglected. As an alternative, $\gamma$-jet production is also simulated using HERWIG7 with the MMHT2014lo~\cite{Harland_Lang_2015} PDF set. Simulation is done separately for samples with fragmentation photons and prompt photons.

The magnitude of the difference between the shape of simulated jet spectra and jet spectra measured in 5.02 TeV $pp$ collisions~\cite{ATLAS:2018gwx} is smaller than approximately 20\% for both the inclusive jets and $\gamma$-tagged jets in both PYTHIA8 and HERWIG7. The jet $\pt$ spectra for inclusive jet samples are reweighted before parameterization (Equation~(\ref{eq:unmod})) to reduce this difference. The reweighting factors are obtained by fitting the data-to-MC ratio by a fifth-order and fourth-order polynomial in $\ptjet$ in the case of PYTHIA8 and HERWIG7, respectively. The same reweighting is applied on quark-initiated and gluon-initiated jets. In the case of $\gamma$-tagged jets, a two-step reweighting procedure is applied. First, the prompt photon and fragmentation photon contributions are scaled by constant weights such that the $\chi^2$ defined using the difference between the $pp$ data and MC is minimized. To further improve the data-to-MC ratio, a third-order polynomial in $\ptjet$ is used to reweight jet spectra connected with prompt photons. 
Alternatively, the reweighting is applied to jet spectra connected with fragmentation photons. 
In general, the differences in the resulting $\Raa$ between reweighted and original spectra are rather small and are quantified later in the paper. 

\begin{figure}
\begin{center}
\includegraphics[width=0.48\textwidth]{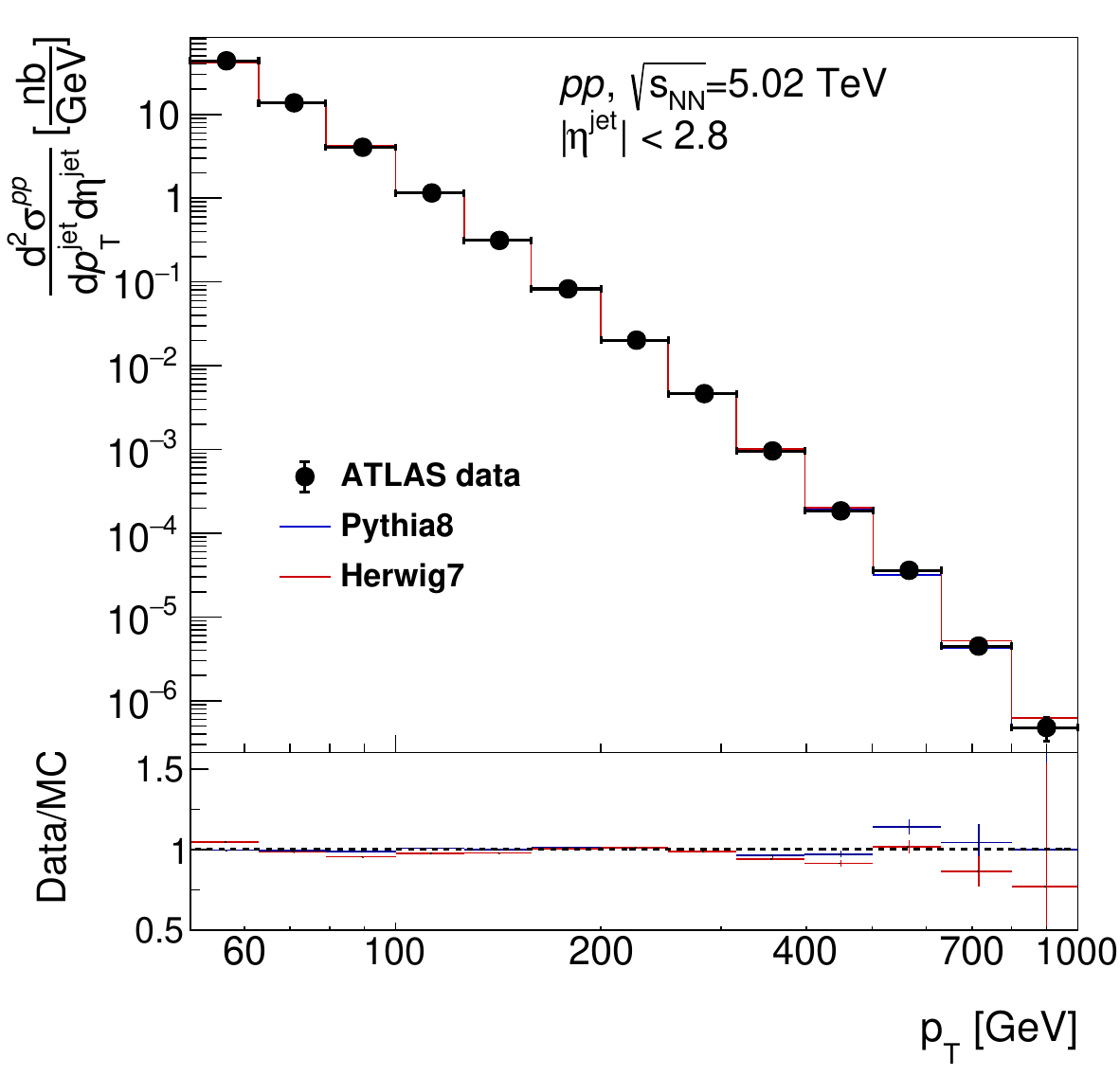}
\includegraphics[width=0.48\textwidth]{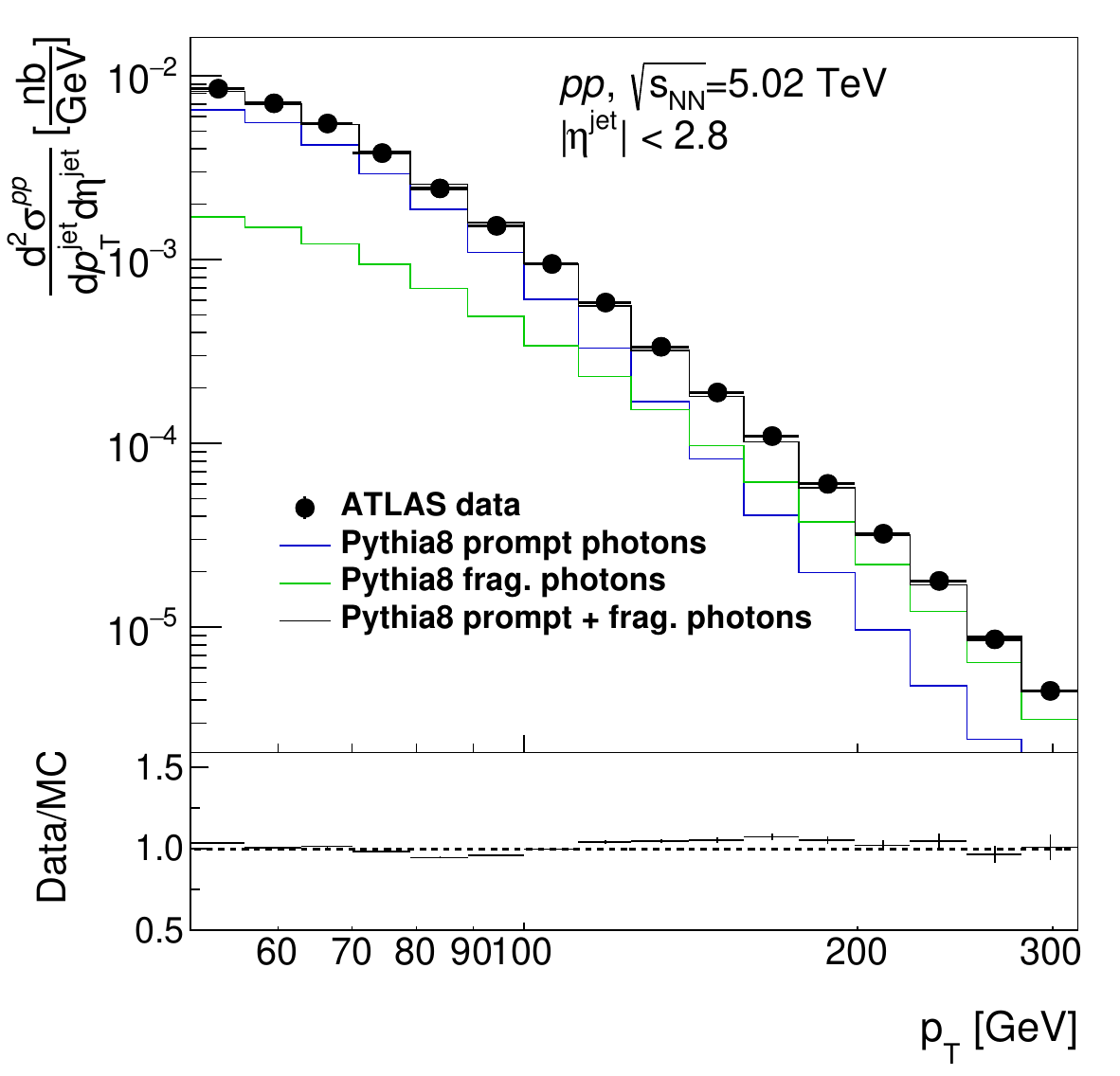}
\end{center}
\caption{\label{fig:spec1} \textit{Left:} Cross-section for the inclusive jet production from PYTHIA8 and HERWIG7 compared with the data \cite{ATLAS:2018gwx}. \textit{Right:} Cross-section for the jet production in $\gamma$-jet system from PYTHIA8 along with individual contributions from prompt and fragmentation photons compared with the data \cite{ATLAS:2023iad}. The reweighting discussed in the text is applied on both MC cross-sections. The ratio of data to MC is shown in the lower panels. }
\end{figure}

The comparison of the parameterization of the final reweighted MC spectra to the 5.02 TeV $pp$ data is shown in Figure~\ref{fig:spec1}. The left panel of Figure~\ref{fig:spec1} shows $\ptjet$ spectra for inclusive jets, and the right panel shows spectra for jets from $\gamma$-jet events. 
One may see that parameterizations of jet $\pt$ spectra are consistent with the data within statistical uncertainties. These parameterizations are used to evaluate the denominator of \Raa.

To simulate the initial state effects in lead-lead collisions, additional samples for both the inclusive jet production and $\gamma$-jet production in PYTHIA8 are produced using a correction for the isospin effect and EPPS16 NLO nuclear parton distribution functions (nPDF) \cite{Eskola:2016oht}. Jet spectra with nPDF and isospin effects are reweighted by factors discussed above, and they are then parameterized using Equation~(\ref{eq:unmod}). Unless stated otherwise, these parameterizations are used in the numerator of \Raa. The alternative sets of nPDFs were also tested, namely EPPS21~\cite{Eskola:2021nhw} and nNNPDF3.0~\cite{AbdulKhalek:2022fyi}, with the conclusion that within the statistical precision of our samples (25 million jet events in total), no differences in derived inclusive jet \Raa were observed. 

All parameterizations of input jet spectra, more details on the impact of nPDFs on jet spectra, and comparison of modeling of quark/gluon fraction in different MC generators are provided in Appendix~\ref{sec:appA}.

\section{Suppression of inclusive jets}
\label{sec:incl}

To determine the free parameters for various jet quenching parameterizations discussed in Section~\ref{sec:mod}, the inclusive jet \Raa measured by ATLAS is used~\cite{ATLAS:2018gwx}. 
The parameters are extracted by a multidimensional minimization procedure based on Minuit \cite{James:1975dr} similar to that used in \cite{Spousta:2016agr}. The ability of a given parameterization to describe jet \Raa data is quantified in terms of $\chi^2/$ndof evaluated as a difference between the model and the data for the $\Raa$ measured differentially in $\ptjet$ and centrality.

First, it is found that jet \Raa from \cite{ATLAS:2018gwx} measured inclusively in rapidity is unable to constrain $c_F$. As a default value of $c_F$ we therefore use $c_F=1.78$, which was found in the previous study using Run-1 data measured differentially in rapidity \cite{Spousta:2016agr}. The value determined in that study $(c_F=1.78 \pm 0.12)$ is consistent with the value of effective color charge calculated and measured for the vacuum jets with hardness $Q \approx 100$~GeV \cite{Capella:1999ms,Acosta:2004js} and it is also consistent with the results of the analysis of Casimir scaling suggesting $c_F \sim 1.7$ \cite{Apolinario:2020nyw}. As alternative values of $c_F$, we use the value of leading-order approximation, $c_F = C_A/C_F=9/4$, and the value used in Hybrid model calculations, $c_F = (9/4)^{1/3} \doteq 1.31$, \cite{Casalderrey-Solana:2014bpa}. 

\begin{figure}
\begin{center}
\includegraphics[width=0.45\textwidth]{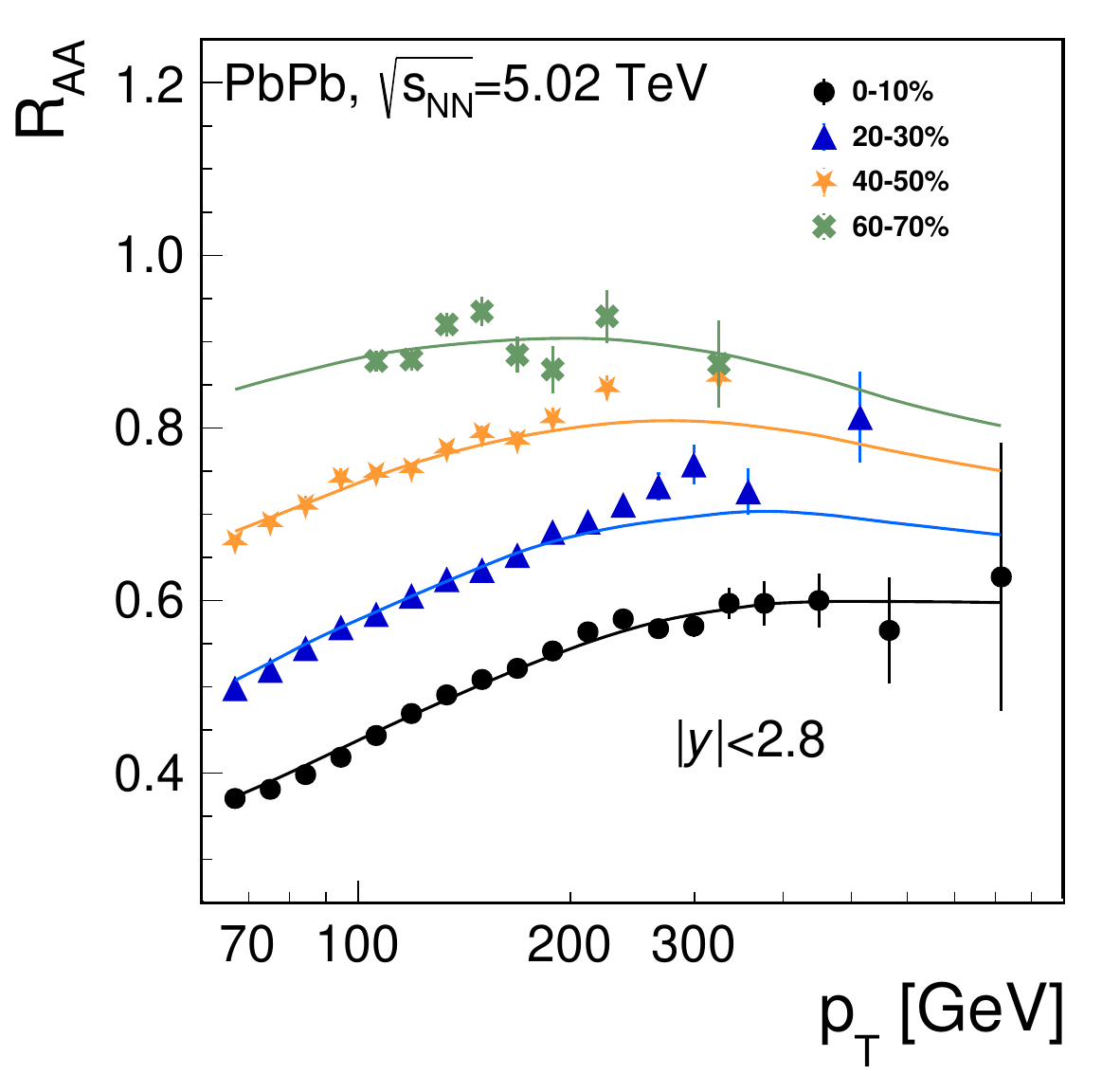}
\includegraphics[width=0.45\textwidth]{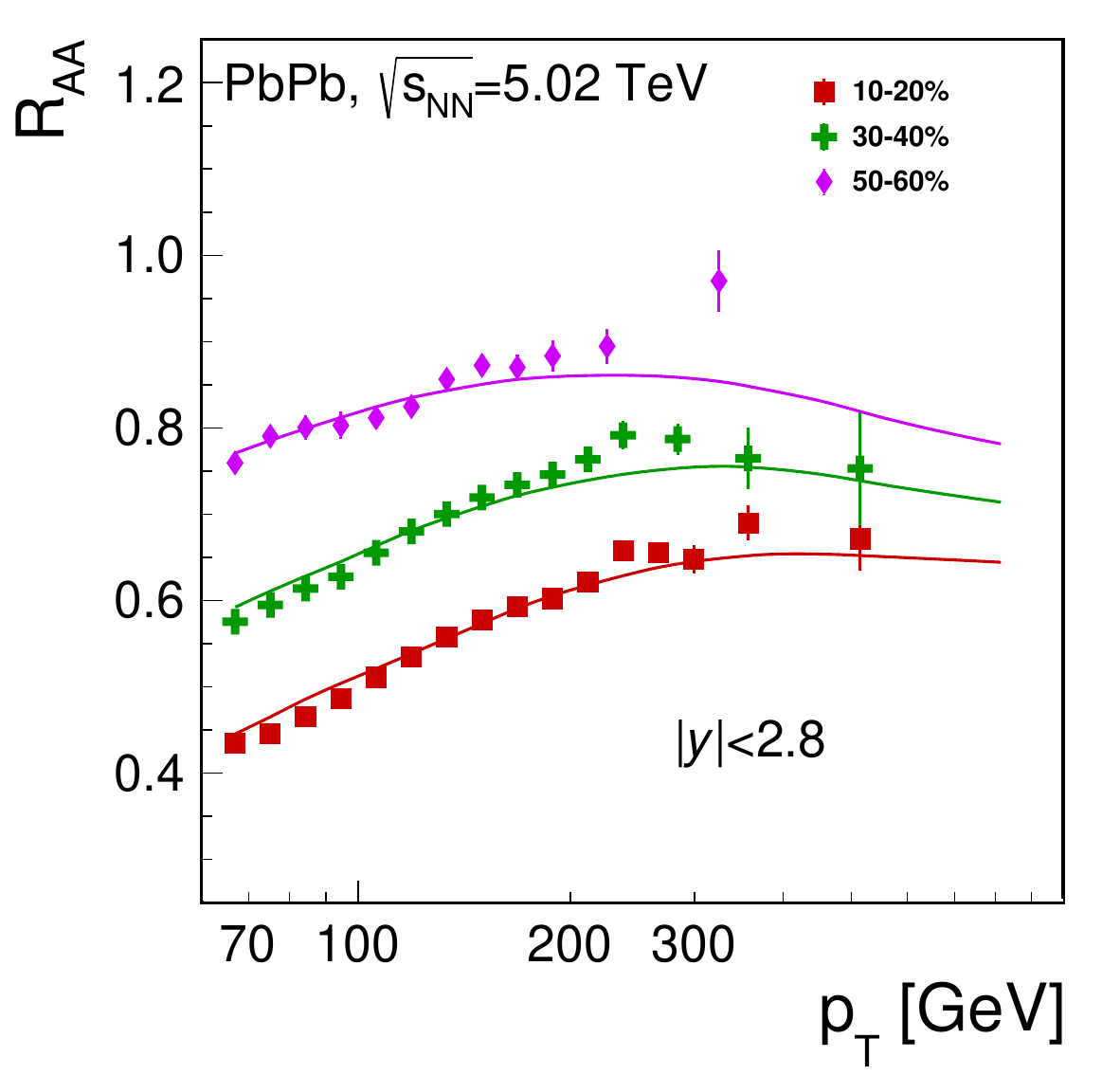} 
\end{center}
\caption{\label{fig:incl} Measured jet \Raa\ \cite{ATLAS:2018gwx} (data points) compared to the parameterization p1 of jet quenching (solid line), which provides the best description of the data (for details on parameterization p1, see the text).}
\end{figure}

The default parameterization, which is found to describe the data with the highest accuracy, uses power-law ansatz for the energy loss from Equation~(\ref{eq:s}), simulation of isospin and nPDFs effects (discussed in Section~\ref{sec:spec}), and parameterization of energy loss fluctuations based on Equation~(\ref{eq:fluct3}) with parameters $c_0 \neq c_1$ (discussed in Section~\ref{sec:mod}).
The comparison of the data and the parameterization is shown in Fig.~\ref{fig:incl}. The power-law parameter is found to be $\alpha=0.27 \pm 0.03$, $\pars_{0-10\%}$ and $\pars_{60-70\%}$ at $\ptjet=100$~GeV were found to be $15.0 \pm 0.3$~GeV and $1.9 \pm 0.1$~GeV, respectively. 

The default parameterization (labeled 'p1') is compared to other parameterizations as summarized in Table~\ref{tab:par}. Parameterizations p2 and p3 represent alternatives to p1 with $c_F=(9/4)^{1/3}$ and $c_F=9/4$, respectively. One may see from the table that three parameterizations, p1, p2, and p3, achieve the same precision. This level of precision can be compared with parameterizations p4 and p5, which do not implement nPDF and isospin effects and which are based on PYTHIA8 and HERWIG7, respectively. One may see that the presence of nPDFs improves the description in the most central collisions, where the highest $\pt$ values are reached and where nPDF effects lead to a flattening of the $\Raa$ as also shown in previous studies (see e.g. Refs.~\cite{Adhya:2021kws,Pablos:2019ngg}). The impact of fluctuations is quantified by parameterizations p6 and p7. In parameterization p6, only the average energy loss is applied. A significant decrease in the ability to describe the data is seen, providing a model-independent argument for the importance of fluctuations in the description of realistic jet quenching. In parameterization p7, an alternative description of fluctuations in jet quenching is used with $c_0=c_1=4$ in (\ref{eq:fluct3}), which implements a somewhat broader spectrum of fluctuations compared to the default one and which reduces the number of parameters connected with the description of fluctuations to just one. This configuration is also outperformed by the default one, though not that significantly. A worse description of the data is also achieved when using a modified power-law ansatz for the energy loss with logarithmic $\pt$ dependence from Equation~(\ref{eq:s2}), which is done in parameterization p8. 
While the agreement between a given parameterization and the data may be further improved for all parameterizations by introducing centrality-dependent parameter $\alpha$, we may conclude that the ansatz with power-law with one centrality-independent parameter in the exponent provides a sufficient description of inclusive jet $\Raa$ (see Figure~\ref{fig:incl} again). The values of all parameters extracted from this study are provided in Appendix~\ref{sec:appB} along with the discussion of the correlations of parameters $\alpha$ and $s$.

\begin{table}
    \centering
    \begin{tabular}{c|c|c|c|c|c|c}
    Label & Formulae & Spectra & Parameters & References & $\chi^2|_{0-10\%}$ & $\chi^2|_\mathrm{all}$ \\
      p1 & (\ref{eq:s}),(\ref{eq:fluct3}) & P8, nPDF & $\alpha_\mr{min}=0.27, \cf=1.78$, $c_0 \neq c_1$ & 
                        \cite{Spousta:2016agr,Apolinario:2020nyw} & 0.51 & 1.06 \\
      p2  & (\ref{eq:s}),(\ref{eq:fluct3}) & P8, nPDF & $\alpha_\mr{min}=0.24, \cf=(9/4)^{1/3}$, $c_0 \neq c_1$ & 
                        \cite{Spousta:2016agr,Apolinario:2020nyw,Casalderrey-Solana:2014bpa} 
                                                                  & 0.53 & 1.05\\
      p3  & (\ref{eq:s}),(\ref{eq:fluct3}) & P8, nPDF & $\alpha_\mr{min}=0.29, \cf=9/4$, $c_0 \neq c_1$ & 
                        \cite{Spousta:2016agr,Apolinario:2020nyw} & 0.50 & 1.09\\
      p4 & (\ref{eq:s}),(\ref{eq:fluct3}) & P8 & $\alpha_\mr{min}=0.33, \cf=1.78$, $c_0 \neq c_1$ & 
                        \cite{Spousta:2016agr,Apolinario:2020nyw} & 0.70 & 1.06 \\                
      p5 & (\ref{eq:s}),(\ref{eq:fluct3}) & H7 & $\alpha_\mr{min}=0.30, \cf=1.78$, $c_0 \neq c_1$ & 
                        \cite{Spousta:2016agr,Apolinario:2020nyw} & 0.88 & 1.18 \\                
      p6 & (\ref{eq:s}) & P8, nPDF & $\alpha_\mr{min}=0.40, \cf=1.78$ & 
                        \cite{Spousta:2016agr}                    & 0.62 & 1.53 \\
      p7 & (\ref{eq:s2}),(\ref{eq:fluct3}) & P8, nPDF & $\alpha_\mr{min}=0.34, \cf=1.78$, $c_0=c_1=4$ & 
                        \cite{He:2018xjv}                         & 0.80 & 1.26 \\
      p8 & (\ref{eq:s2}),(\ref{eq:fluct3}) & P8, nPDF & $\alpha_\mr{min}=0.15, \cf=1.78$, $c_0=c_1=5$ & 
                        \cite{He:2018xjv}                         & 0.44 & 1.43 \\
    \end{tabular}
    \caption{Definitions of parameterizations of $\ptjet$ dependence of jet quenching along with parameter $\alpha$ obtained by minimization with respect to measured jet $\Raa$~\cite{ATLAS:2018gwx}. The precision of the description of measured jet $\Raa$ by a given parameterization is quantified in terms of $\chi^2$ evaluated for $0-10\%$ centrality bin and for all centrality bins.}
    \label{tab:par}
\end{table}

Equipped with the parameterization, we will proceed with estimates of path-length dependence of the parton energy loss and provide a simple extrapolation of jet quenching magnitude from lead-lead to oxygen-oxygen collisions.

\section{Path length dependence and system size dependence of jet suppression}
\label{sec:path}

\subsection{Path length dependence and jet \Raa}

The path-length ($L$) dependence of the jet energy loss may be quantified as $\pars \sim \parl^\delta$ with $\delta=1,2,$ and 3 employed in baseline calculations of collisional energy loss, pQCD radiative energy loss, and non-perturbative strong coupling models, respectively \cite{Majumder:2010qh,Baier:1996kr,Gubser:2008as}. 
During the past ten years, several studies were done aiming to extract the path-length dependence of parton energy loss from the data on charged particle \Raa \cite{Betz:2014cza,Betz:2016ayq,Djordjevic:2018ita,Arleo:2022shs}. Large comparative studies \cite{Betz:2014cza,Betz:2016ayq} which also used data on the elliptic flow led to inconclusive results of $ 1 \leq \delta \leq 3$. More recent studies \cite{Djordjevic:2018ita} and \cite{Arleo:2022shs} led to $\delta = 1.4$ and $\delta \sim 1.02$ at high-\pt, respectively. These studies rely on various assumptions such as modeling of medium expansion (which was shown to have an impact on observables directly depending on path-length \cite{Adhya:2021kws}), power-law approximation of input parton yields (which was shown to be inaccurate \cite{Spousta:2015fca}), or relationship between energy loss of parton shower and measured charged hadron yields. Studies of path-length dependence of parton energy loss based on jet \Raa were also performed based on LBT model of jet quenching \cite{Wu:2023azi}, finding $\delta < 1$.  

To extract the path-length dependence of the energy loss, we use parameterizations described in the previous section, which present results on jet quenching modeling with minimal assumptions on the actual jet quenching mechanism. To model the geometry, we use two approaches. In the first approach, we assume that the average path length is proportional to the path length estimates given by the Glauber model from Ref.~\cite{Loizides:2017ack}. This represents the simplest possible assumption, which is, however, at least partially validated by the measurements of electro-weak boson production \cite{ATLAS:2015rlt,CMS:2020oen,ATLAS:2019ibd,CMS:2012fgk,ATLAS:2019maq,CMS:2014dyj}, confirming good predictive power of the Glauber model. In what follows, this approach is labeled {\it Glauber}. The second approach to the modeling of geometry uses two estimates of the average path length based on calculations with Trajectum \cite{Nijs:2020ors,Nijs:2020roc} taken from Ref.~\cite{Beattie:2022ojg}. In those calculations, the full hydrodynamical simulation of medium expansion is performed with the initial conditions obtained from the TRENTO model \cite{Moreland:2014oya}. 
The first estimate is based on integrating the path along lines connecting the nucleon-nucleon collision with the freeze-out hypersurface. The second estimate uses the same approach as the first one, but it takes into account the local fluid velocity $u_{\mu}$ and overall boost of the fluid quantified by Lorentz factor $\gamma$. The average path length is then calculated by integrating with the weight, $u_{\mu}/\gamma$. In both estimates, the integration starts at $\tau=0.37$~fm/$c$ since before this time the hydrodynamical evolution is not well defined. In what follows, the former and latter estimates are labeled for simplicity as {\it Trajectum1} and {\it Trajectum2}, respectively. In Fig.~2 of Ref.~\cite{Beattie:2022ojg}, which provides estimates used here, they are labeled by $L_\mathrm{dyn}$ and $\int u_{\mu}/\gamma \mathrm{d}L^{\mu}$, respectively.

\begin{figure}
\begin{center}
\includegraphics[width=0.45\textwidth]{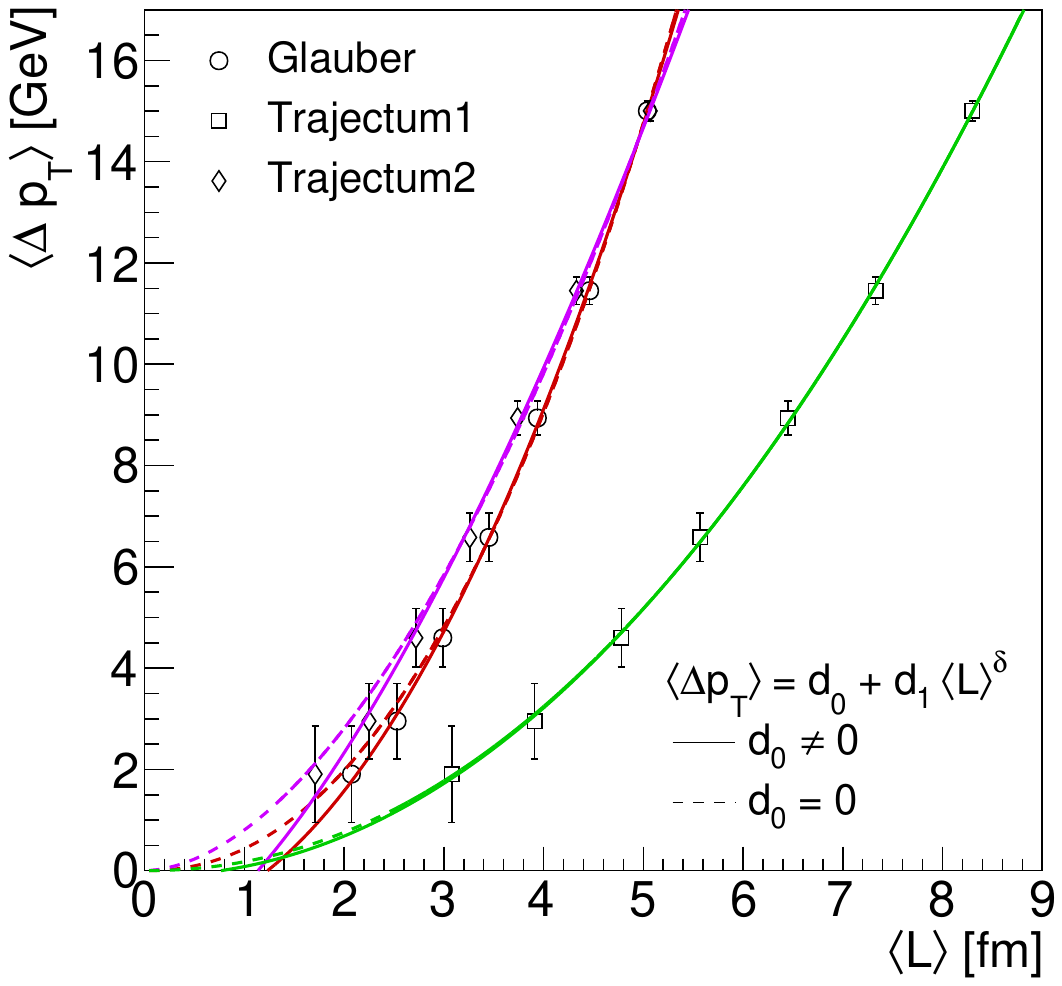}
\includegraphics[width=0.45\textwidth]{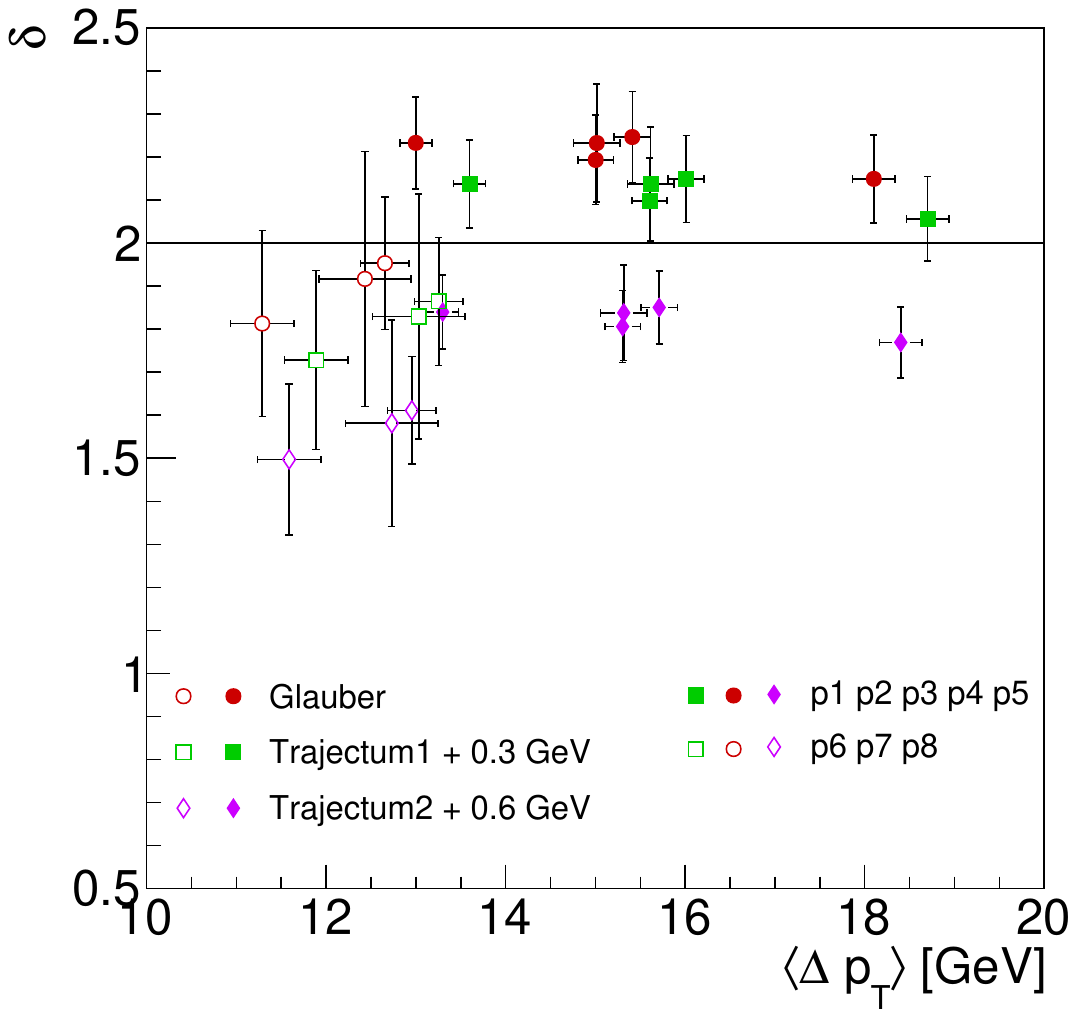} 
\end{center}
\caption{\label{fig:path} 
\textit{Left:} Path-length dependence of average jet energy loss extracted from the jet $\Raa$ data using the parametric model for parameterization p1 and path lengths calculated using Glauber and Trajectum. Solid and dashed lines show fits with $d_0 \neq 0$ and $d_0=0$, respectively. 
\textit{Right:} Power $\delta$ of the path-length dependence of jet energy loss ($y$ axis) versus average lost energy ($x$ axis) obtained for parameterizations of energy loss p1-p5 (closed markers) and p6-p8 (open markers). Path lengths are calculated using Glauber and Trajectum. Path lengths from Trajectum are taken from Ref.~\cite{Beattie:2022ojg}. Values for Trajectum1 and Trajectum2 are shifted by 0.3~GeV and 0.6~GeV along the $x$-axis for better legibility. For details, see the text.}
\end{figure}

The left panel of Figure~\ref{fig:path} shows the fit by power-law parameterization of path length,
\begin{equation}
\label{eq:L}
   \pars = d_0 + d_1 \parl^\delta,   
\end{equation}
with $d_0, d_1,$ and $\delta$ being parameters of the fit. The fit is performed for $\pars$ evaluated as a function of $\parl$ for p1 parameterization. The $\pars$ is evaluated for the initial jet $\pt=100$~GeV. Two versions of the fit are done, one with $d_0$ free and one with $d_0$ fixed to $0$. The error bars on the data points come from propagation of both the bin-wise uncorrelated and fully correlated uncertainties in the jet $\Raa$ measurement, where the latter represent the dominant source of the uncertainties. One can see a good description of the distributions by the fits for all three models defining $\parl$. The values of $\delta$ parameters for Glauber, Trajectum1, and Trajectum2 are 
$2.0 \pm 0.4, 2.1 \pm 0.3$, and $1.5 \pm 0.3$, respectively. For Glauber and Trajectum1, the value of $\delta$ is consistent with two. Trajectum2 leads to a lower value of $\delta$, which may be understood as a consequence of the effective reduction of the path length due to taking into account the effect of the co-moving medium. The values of $d_0$ parameters for Glauber, Trajectum1, and Trajectum2 are
$-1.0 \pm 1.8, -0.1 \pm 1.5$, and $-1.7 \pm 1.9$ GeV.  For all three path-length models, the value of $d_0$ is negative but consistent with zero. Negative $d_0$ may be connected with the fact that jets do not start losing energy exactly at $\tau=0$, see e.g. discussions in Ref.~\cite{Andres:2019eus,Adhya:2019qse,Caucal:2018dla}. The non-zero $d_0$ may also be connected with a limited precision of the jet $\Raa$ measurement in peripheral collisions or the limited precision of the modeling. Consequently, we also perform the fit with $d_0$ fixed to zero for a comparison. The resulting values of $\delta$ parameters for Glauber, Trajectum1, and Trajectum2 are $2.2 \pm 0.1, 2.1 \pm 0.1$, and $1.8 \pm 0.1$, respectively.  

The fit results for parameter $\delta$ evaluated as a function of $\pars$ are summarized in the right panel of Figure~\ref{fig:path} for all parameterizations of jet quenching from Section~\ref{sec:incl} and for the fit with $d_0$ fixed to zero. One can see from the figure that different jet quenching scenarios provide rather different values of $\pars$, and thus different values of jet quenching parameter $\hat{q}$, which we do not extract from the data. At the same time, the value of $\delta$ is stable for all parameterizations p1-p5, and it is close to the value of $2$ for all three path-length models. For parameterizations p6-p8, which exhibited higher $\chi^2$ in comparison with the data, the values are in general below the value of 2, albeit with larger uncertainties, which is connected with a worse description of the data in more peripheral collisions. The average values of $\delta$, $d_0$, $d_1$, and $L_0$ calculated as $L(\pars=0)$ are provided in Table~\ref{tab:len}. The values are calculated as averages for parameterizations p1-p5 for each of the path-length models. One can see from the table that $\delta$ values are broadly consistent with 2 for Glauber and Trajectum1, while for Trajectum2, they are significantly smaller. In general, these results may be interpreted as supporting the radiative nature of the parton energy loss based on parametric modeling with a small number of input parameters. At the same time, these results also highlight the importance of modeling the dynamical aspects of energy loss in a co-moving, expanding medium.

The results from this section and Section~\ref{sec:incl} are further used to evaluate the dependence of parton energy loss on the number of participating nucleons ($\npart$) estimated by the Glauber model, which is done in Appendix~\ref{sec:appC}. 

\begin{table}
    \centering
    \begin{tabular}{c|c|c|c|}
    parameters & Glauber  &  Trajectum1  &  Trajectum2 \\ 
    
    $\langle \delta \rangle ~ (d_{0} =    0)$ &$  2.21 \pm 0.05 $&$  2.12 \pm 0.05  $&$  1.82 \pm 0.04 $ \\ 
    $\langle \delta \rangle ~ (d_{0} \neq 0)$ &$  2.01 \pm 0.17 $&$  2.10 \pm 0.16  $&$  1.55 \pm 0.13 $ \\ 
    $\langle d_{1} \rangle  ~ (d_{0} =    0)$ &$  0.43 \pm 0.03 $&$  0.18 \pm 0.02  $&$  0.80 \pm 0.05 $ \\ 
    $\langle d_{1} \rangle  ~ (d_{0} \neq 0)$ &$  0.64 \pm 0.22 $&$  0.19 \pm 0.08  $&$  1.39 \pm 0.38 $ \\ 
    $\langle d_{0} \rangle$ [GeV]             &$  -1.0 \pm 0.9 $&$  -0.1 \pm 0.7  $&$  -1.7 \pm 0.9 $ \\ 
    $\langle L_{0} \rangle$ [fm]              &$  1.2 \pm 0.6 $&$  1.1 \pm 0.5  $&$  0.0 \pm 0.0 $ \\ 

    \end{tabular}
    \caption{Average values of parameters from the fit of path-length dependence of energy loss for path lengths calculated using Glauber and Trajectum. Path lengths from Trajectum are taken from Ref.~\cite{Beattie:2022ojg}. Values are obtained as averages from five energy loss parameterizations, p1-p5.}
    \label{tab:len}
\end{table}

\subsection{Path length dependence and jet $v_2$}

Given the path-length parameterization of energy loss, one may continue exploring the path-length dependence of jet quenching by evaluating jet $v_2$. Jet $v_2$ quantifies the magnitude of azimuthal-angle modulation of jet yields, which is due to the difference between the energy loss of jets traveling in the direction of the interaction plane (``in'' direction) and jets traveling in the direction perpendicular to the interaction plane (``out'' direction). The former suffer the energy loss over the path-length $L_{in}$, while the latter suffer the energy loss over the path-length $L_{out}$. The jet $v_2$ may then be estimated as \cite{Zigic:2018smz}
  ~
\begin{equation}
    v_2(\ptjet) \approx \frac{1}{2} \frac{\Raa(\ptjet,L_{in}) - \Raa(\ptjet,L_{out})}{\Raa(\ptjet,L_{in}) + \Raa(\ptjet,L_{out})}.
    \label{eq:v2}
\end{equation}
  ~
To calculate $v_2$, one needs to use estimates of the average path length traveled in the directions in and out, which differ from the average path length, $\parl$, by factors $\Delta L_{in}$ and $\Delta L_{out}$, respectively. The $L_{in}$ and $L_{out}$ in Equation~(\ref{eq:v2}) are then given by
\begin{eqnarray}
    \nonumber
    L_{in} = \parl - c_l \cdot \Delta L_{in}, \\
    L_{out} = \parl + c_l \cdot \Delta L_{out}.
    \label{eq:v2L}
\end{eqnarray}
Since the relative difference between $\parl$ and $L_{in}$ or $L_{out}$ may differ in reality from the estimates by the path-length models, a free parameter $c_l$ is introduced, which is, however, enforced to be the same for all the centrality bins in this analysis. The $c_l=1$ would imply traveling through the medium of the same shape as that given by the model. The $c_l<1$ implies a smearing of the shape of the medium, $c_l>1$ implies an amplification of the azimuthal differences. 

To obtain values of $\Delta L_{in}$ and $\Delta L_{out}$, we use again the Glauber model and results published in Fig.~5 and Fig.~2 in Ref.~\cite{Beattie:2022ojg} for earlier introduced Trajectum1 and Trajectum2 configurations.  
The comparison between the calculated $v_2$ and data \cite{ATLAS:2021ktw} is shown in the right panel of Figure~\ref{fig:OO}. A good description of the data is found only when applying scaling factors $c_l=0.35, 3.0$, and $0.55$ for Glauber, Trajectum1, and Trajectum2, respectively. This suggests that $\Delta L_{in}$ and $\Delta L_{out}$ in reality differ significantly from estimates based on the models. While results for Glauber and Trajectum2 are similar, results for Trajectum1 differ significantly from these two, which is a direct consequence of the very small $L_{out}/L_{in}$ ratio predicted by the Trajectum1 model (see Fig.~5 left in \cite{Beattie:2022ojg}). The large differences in $c_l$ values for different path-length models clearly demonstrate the large sensitivity of jet $v_2$ observable to details of path-length modeling. The fact that a good description of $v_2(\pt)$ is achieved for all the centrality bins after applying a single scaling factor may imply that the shape of $v_2(\pt)$ and relative magnitude of $v_2$ between different centrality bins is largely dictated by the magnitude of the average energy loss. We should also notice that a good description of the data is not achieved in the $0-5\%$ centrality bin, where the path-length models predict significant jet $v_2$, while the jet $v_2$ measured in the data is consistent with zero. This may be connected with the presence of a sizable $v_3$ component, which is measured to be much larger than $v_2$, and the formula (\ref{eq:v2}) is therefore expected to fail. 

We may now attempt to connect the results presented here with the effective value of $\delta$ parameter obtained from recent path length analyses in 
Refs.~\cite{Arleo:2022shs,Wu:2023azi}. 
When using Equations (\ref{eq:v2L}), one can obtain the following relation for $v_2$ by Taylor expanding $\Raa(\parl + c_l 
\cdot \Delta L)$ around $\parl$, following the same procedure and assumption on the $\pt$ dependence of energy loss as in~\cite{Arleo:2022shs}: 
  ~
\begin{equation}
    v_2(\pt) \approx \frac{1}{2} \frac{\pt}{\Raa} \frac{\partial \Raa}{\partial \pt} \, c_l \, \delta \, e 
\label{eq:arleo}
\end{equation}
where eccentricity $e \approx (\Delta L_{in} + \Delta L_{out})/(2\parl)$. When comparing Equation~(\ref{eq:arleo}) with Equation~(18) in Ref.~\cite{Arleo:2022shs} one can see that they are identical except for the presence of $c_l$ which relates the $\delta$ value obtained in our analysis to the power parameter $\delta'$ obtained in other analyses which do not scale the medium profiles, namely $\delta' = c_l \delta$. When evaluating $\delta'$ using $c_l$ and $\delta$ estimated here, one obtains, $\delta'$ of approximately  0.7 and 0.8 for Glauber and Trajectum2, respectively. This brings the values from our calculations close to the value of the effective power obtained in \cite{Wu:2023azi} and \cite{Arleo:2022shs},  
which are 0.6 and 1.0, respectively. We should stress that our calculations and calculations in \cite{Wu:2023azi} use full jets while calculations in \cite{Arleo:2022shs} use charged hadrons, which is expected to induce a difference in the value of $\delta'$ as discussed already in \cite{Wu:2023azi}. Finally, we should mention that for Trajectum1 we obtain $\delta' \approx 6$, which is an unrealistic value.

Based on the results presented in this section, we can conclude that jet $v_2$ is highly sensitive to differences between in-plane and out-of-plane path length estimates, which may be rather large in the models and which substantially affect the value of the extracted coefficient $\delta$.

\subsection{Jet \Raa in oxygen-oxygen collisions}

\begin{figure}
\begin{center}
\includegraphics[width=0.45\textwidth]{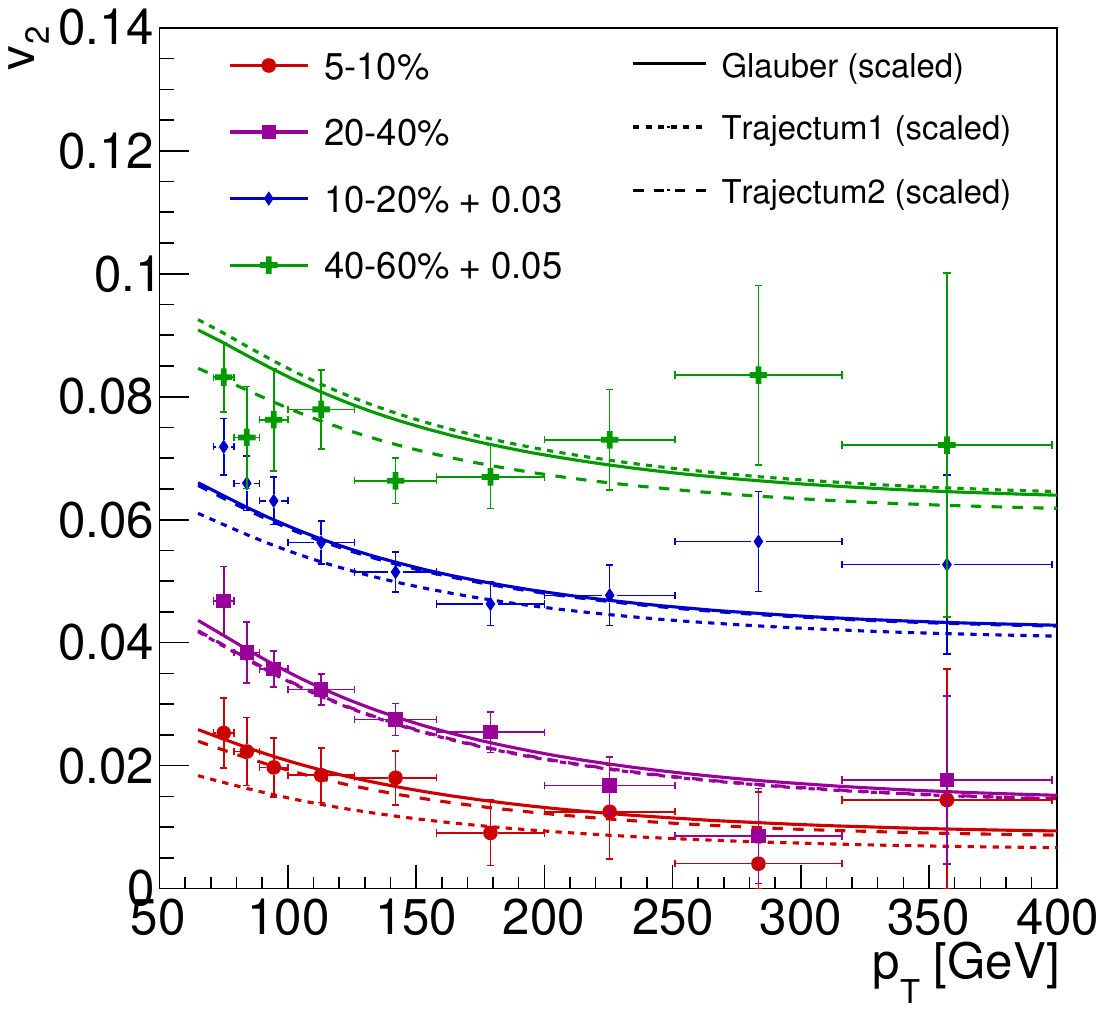}
\includegraphics[width=0.45\textwidth]{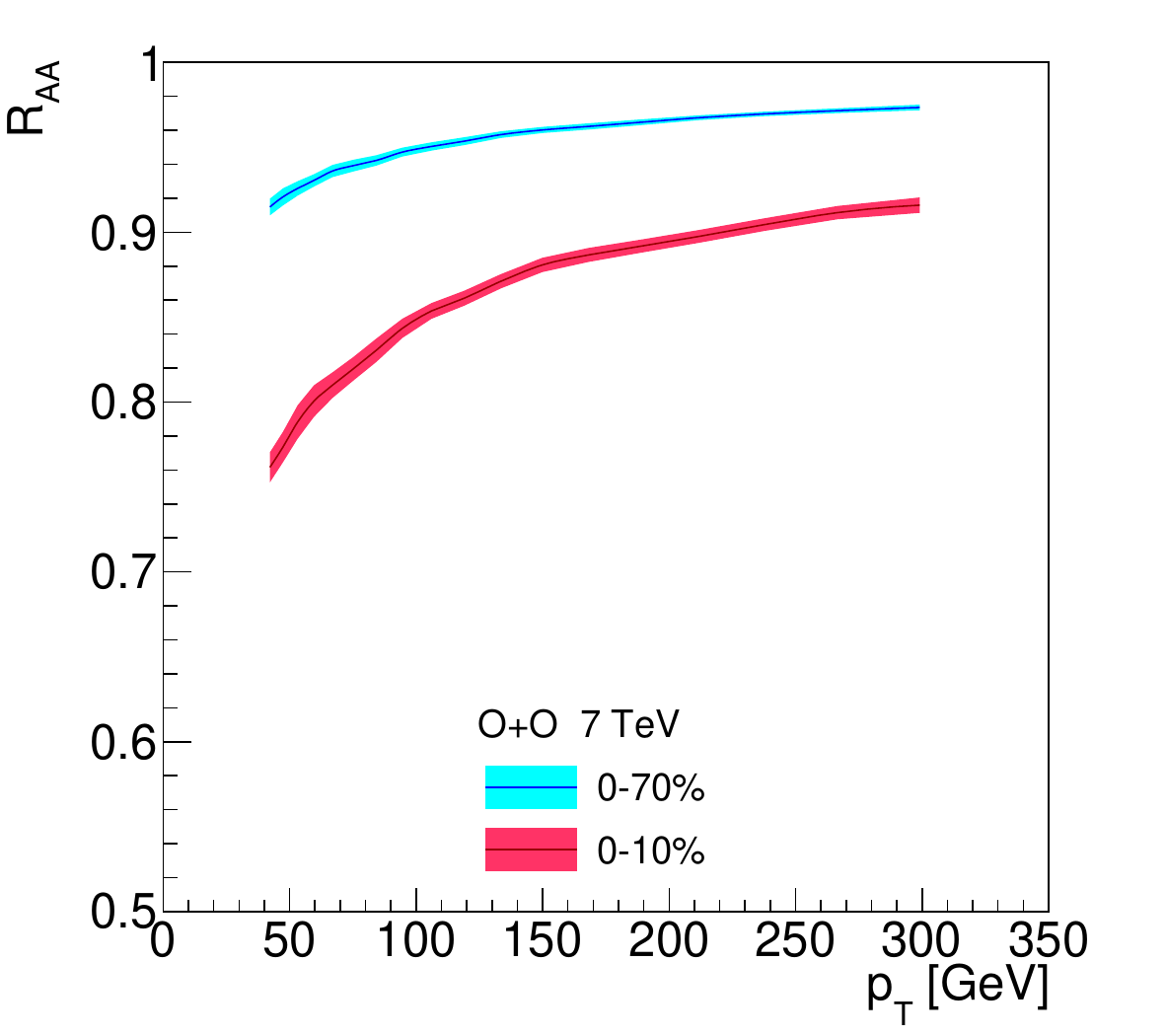}
\end{center}
\caption{\label{fig:OO} \textit{Left:} Jet $v_2$ determined using the parameterization of path-length dependence of inclusive jet suppression (lines) compared with measured jet $v_2$ \cite{ATLAS:2021ktw} (data points). Path lengths are calculated using Glauber and Trajectum and are scaled as described in the text. Path lengths from Trajectum are taken from Ref.~\cite{Beattie:2022ojg}. \textit{Right:} Prediction of jet $\Raa$ distributions for 7 TeV oxygen-oxygen collisions using parameterizations of inclusive jet suppression at 2.76 TeV and 5.02 TeV.}
\end{figure}

We now turn our attention to the system-size dependence of energy loss.
Besides exploring the path-length dependence of parton energy loss, the parameterization of the energy loss may be used to predict the magnitude of jet \Raa in upcoming oxygen-oxygen (O+O) collisions. First, the linear extrapolation of the energy loss to 7~TeV is done from $\pars$ for 2.76~TeV jet spectra determined in Ref.~\cite{Spousta:2016agr} and \pars for 5.02~TeV jets calculated here.  
The value of $\pars$ at 100 GeV is used as a reference value to determine scale factors used to scale the Pb+Pb 5.02 TeV \Raa calculations to O+O 7~TeV \Raa. The scale factors are determined using parameterizations p1-p3, and they vary from 1.25 to 1.42. The average path lengths in two different percentiles of O+O collisions, namely $0-10\%$ and $0-70\%$, are calculated using the Glauber model \cite{Loizides:2017ack} along with centrality analysis published in Ref.~\cite{Behera:2021zhi}. The above-discussed parameterization of $\pars(\parl)$ dependence is then used to determine $\pars$ in O+O collisions for each of p1-p3. Input jet spectra at 7~TeV were obtained from PYTHIA8 with the same settings as described in Section~\ref{sec:spec} but with no simulation of nPDF effects. The resulting prediction is shown for $0-10\%$ and $0-70\%$ O+O collisions in Fig.~\ref{fig:OO}. The band represents the envelope of the results obtained from three parameterizations, p1-p3, and does not attempt to provide a full estimate of uncertainties (for that see e.g. Ref.~\cite{Gebhard:2024flv}). One can see that in $0-10\%$, the suppression is significant with the \Raa of $\approx 0.76$ at $\ptjet = 50$~GeV. In $0-70\%$, the $\Raa$ value is $\approx 0.92$ and 0.98 at $\ptjet = 50$~GeV and $\ptjet = 300$~GeV, respectively. These results represent a baseline prediction for jet quenching in O+O collisions, which may be seen in the data if the underlying energy loss mechanism is the same in O+O and Pb+Pb collisions in the given kinematic window, if the role of fluctuations is comparable, and if nPDF effects have a small impact on the jet $\Raa$. Predictions presented here add new information to detailed predictions for charged hadron $\Raa$ in O+O published in Refs.~\cite{Huss:2020whe,Behera:2023oxe}.

\subsection{Jet $\Raa$ of $b$-jets}

The last topic we briefly discuss is the parton mass dependence of energy loss. To do that, we repeat the minimization from Section~\ref{sec:incl} using the measurement of $b$-jet $\Raa$ from Ref.~\cite{ATLAS:2022agz}. The parameterization of $b$-jet spectra uses the large inclusive jet sample described in Section~\ref{sec:spec}. The extracted magnitude of energy loss is then fitted by the power-law path-length dependence of the energy loss (Equation~(\ref{eq:L})), resulting in parameter $\delta=2.6 \pm 1.6$. This is further compared with the energy loss extracted from the $\Raa$ of $R=0.2$ jets leading to $\delta=1.9 \pm 0.4$. Both values are, therefore, consistent with each other within the uncertainties, and one can conclude that the current precision of $b$-jet measurement does not provide sufficient precision to discriminate expected differences \cite{Zigic:2018ovr,Dokshitzer:2001zm,Wicks:2005gt,Ke:2018tsh} in the path-length dependence of energy loss between inclusive jets and $b$-jets.
For completeness, we should mention that in the fits of both $b$-jets and $R=0.2$ jets, we used the Glauber model only to obtain path-length estimates, and that the fully correlated systematic uncertainties are not taken into account since they do not differ between the two measurements.

To enhance the ability to extract useful information from the $b$-jet measurements in the future, it is especially important to measure the suppression in more centrality bins since, for fewer centrality bins, the precision of extracted $\delta$ significantly decreases.
To quantify the potential improvement in the precision with increasing number of centrality bins, we evaluate the $\delta$ parameter for $R=0.4$ jets using the jet \Raa\ measurement performed in seven centrality bins  \cite{ATLAS:2018gwx}, taking into account only the systematic uncertainties uncorrelated in $\ptjet$. This gives $\delta=2.01 \pm 0.08$ for the Glauber model path lengths. The precision of this value can be compared with the precision of the $\delta$ obtained for $R=0.2$ jets provided above ($\delta=1.9 \pm 0.4$). Given that both measurements have a similar magnitude of uncorrelated systematic uncertainties, but the $R=0.2$ jet measurement is performed only in three centrality bins, one can directly see the impact of performing the measurement in fine bins of centrality from this comparison.

We will now turn our attention to gaining insights on 
the jet suppression in $\gamma$-jet events.

\section{Suppression of jets in $\gamma$-jet system}
\label{sec:gamma}

\begin{figure} [h]
\begin{center}
\includegraphics[width=0.4\textwidth]{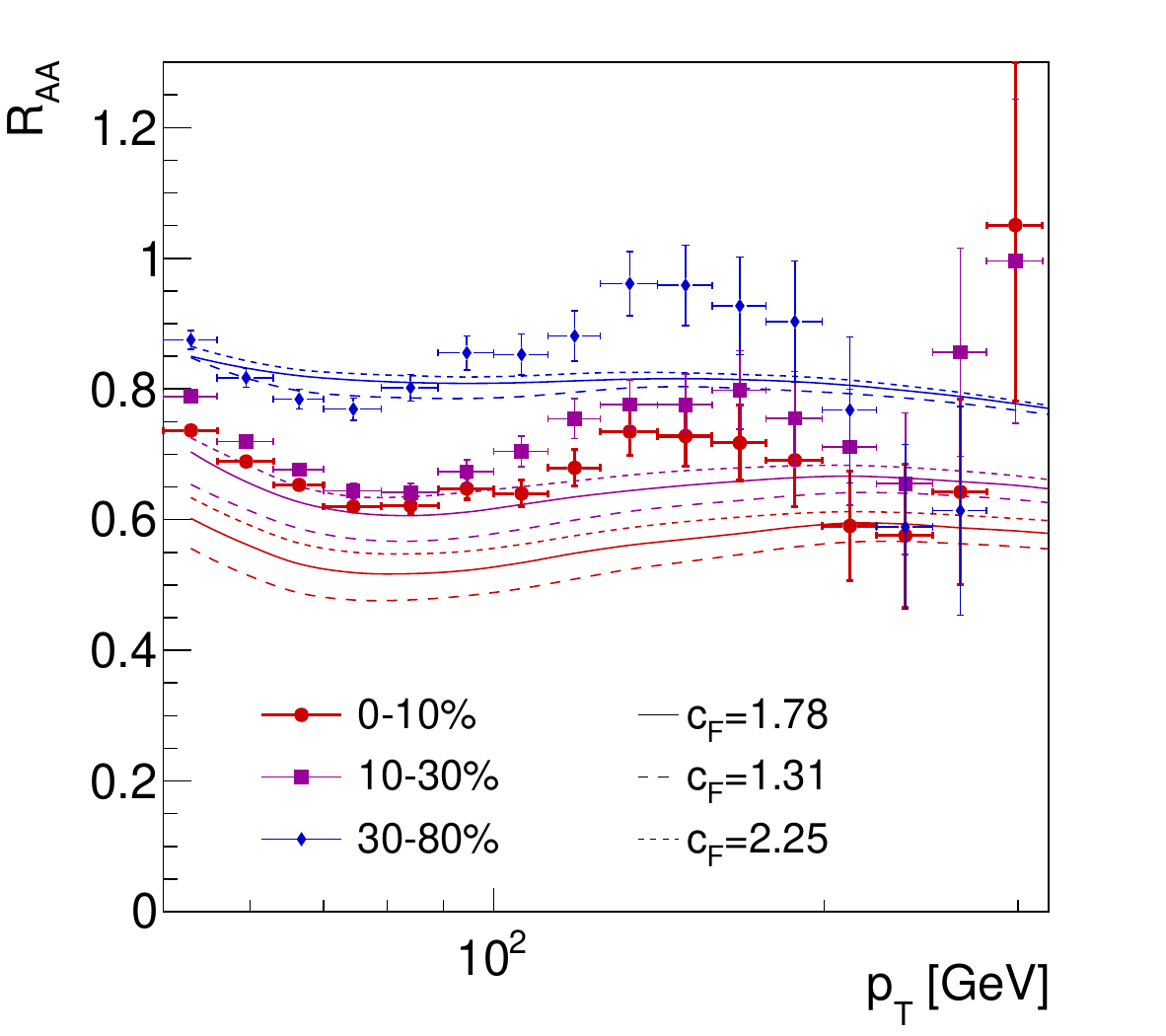} 
\includegraphics[width=0.4\textwidth]{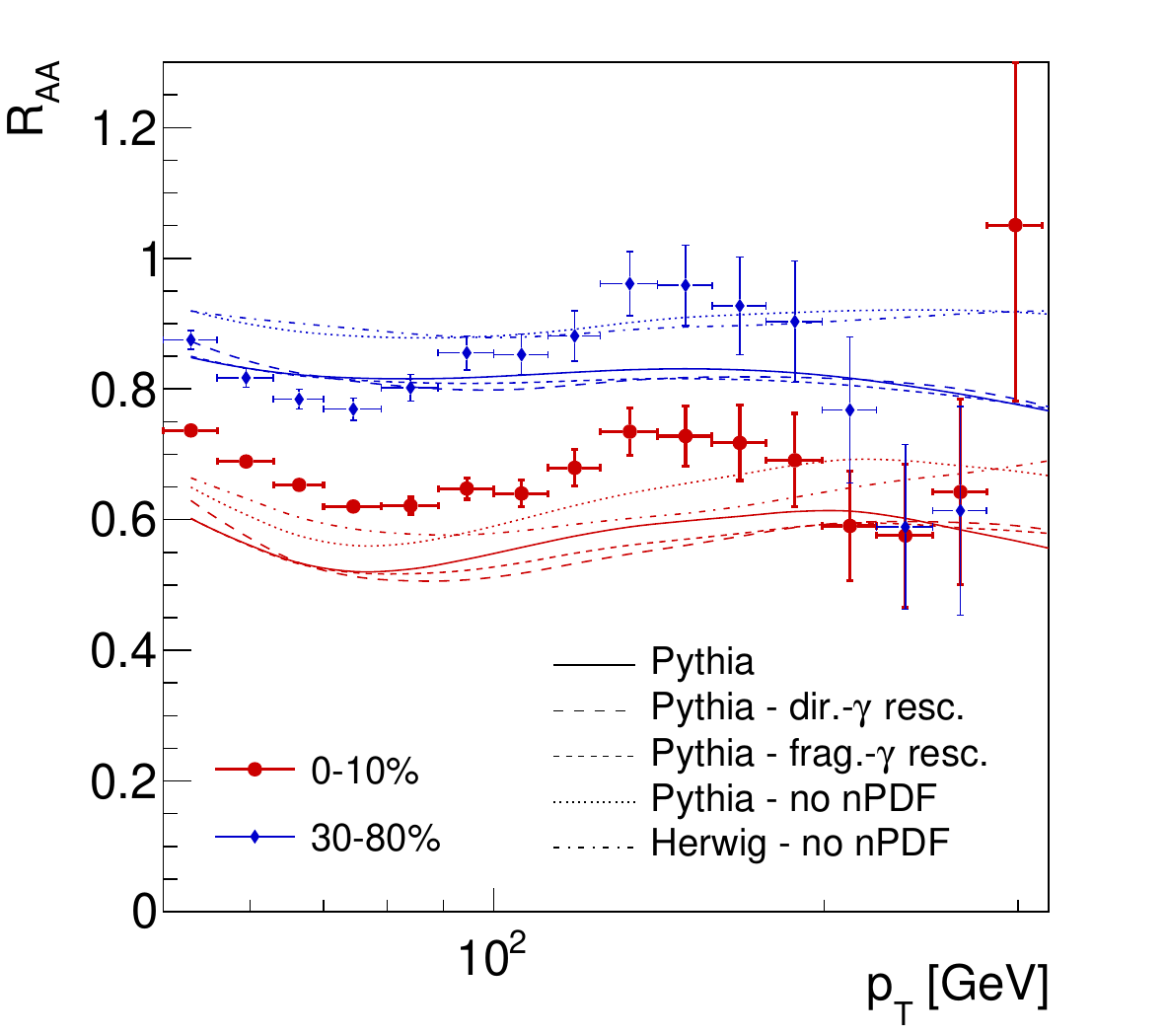} 
\end{center}
\begin{center}
\includegraphics[width=0.4\textwidth]{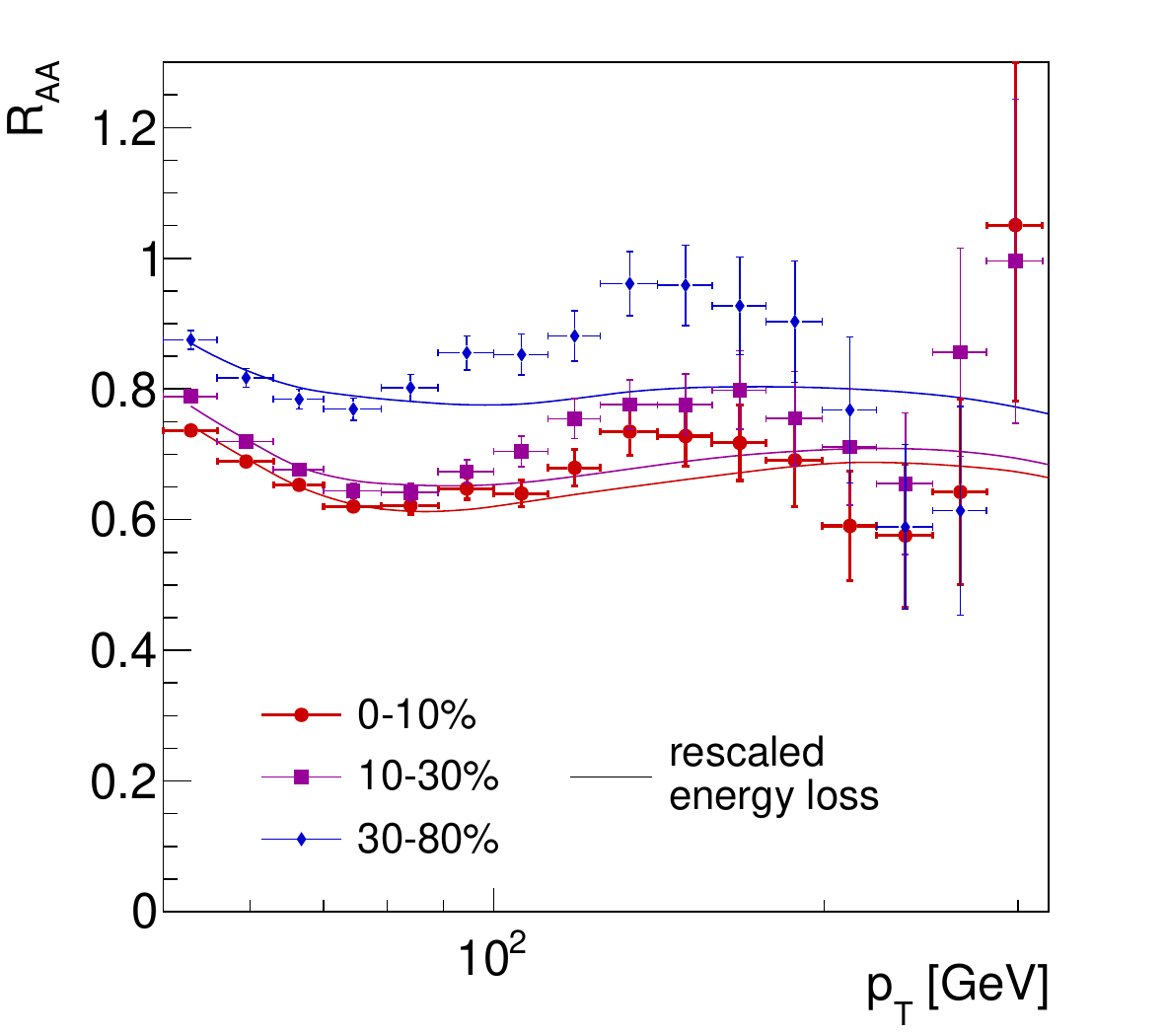}
\includegraphics[width=0.4\textwidth]{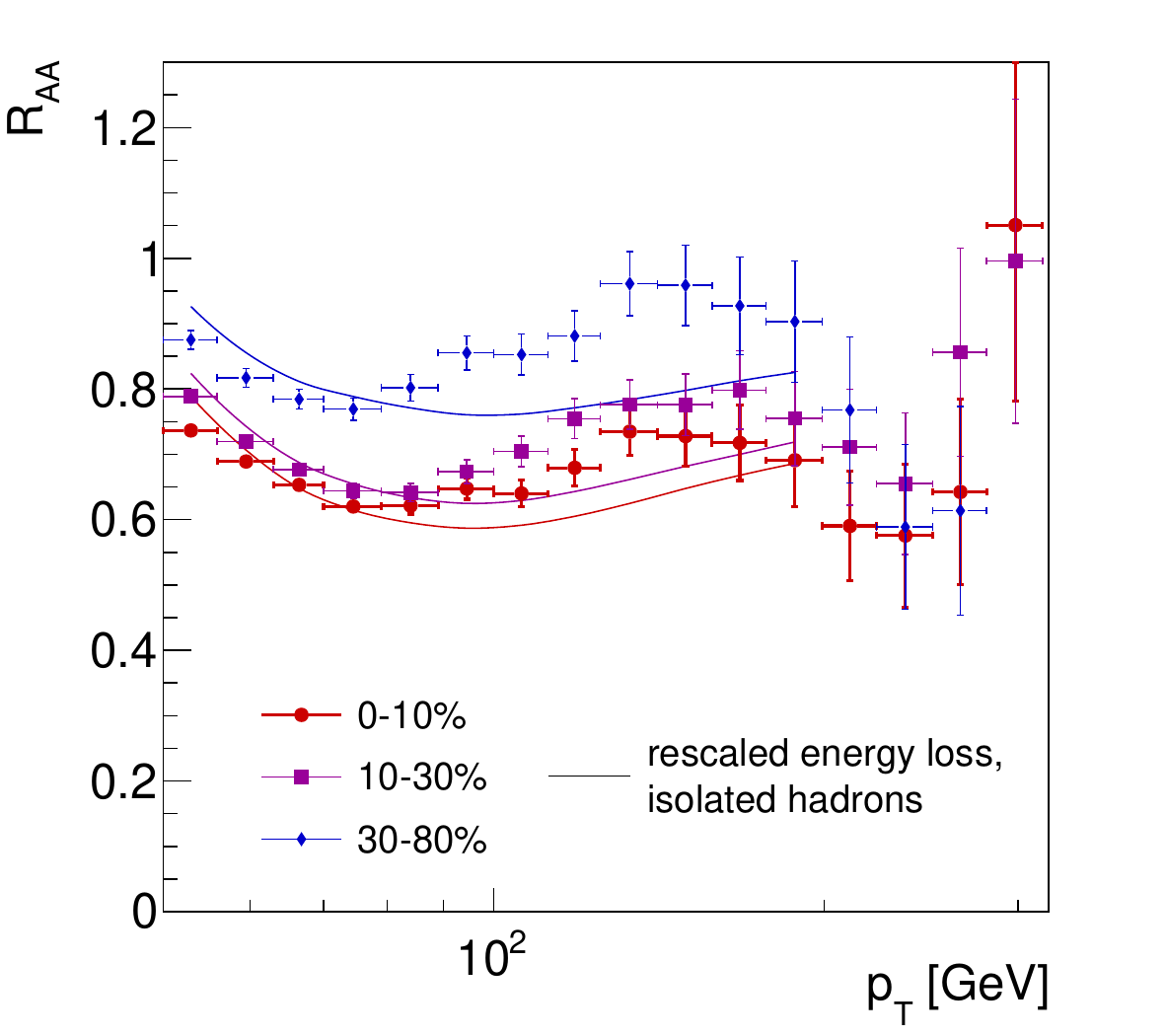}
\end{center}
\caption{\label{fig:gamma} 
  Jet $\Raa$ measured in the $\gamma$-jet system \cite{ATLAS:2023iad} (data points) compared with various setup of the parametric modelling (lines): different choice 
  of color factor (upper left), different choice of input jet spectra (upper right), and rescaled parameterization p1 with input spectra from PYTHIA with rescaled 
  direct photons (lower left). The lower right panel shows jet $\Raa$ evaluated for jets from the isolated-hadron-jet system with rescaled parameterization p1.
}
\end{figure}

Having in hand parameters describing well the suppression of inclusive jets, we may proceed with applying the suppression on jets in the $\gamma$-jet system. First, we use the reweighted PYTHIA8 jet spectra with nPDFs as described in Section~\ref{sec:spec} and calculate the $\Raa$ for the quenching parameterizations p1-p3 which employ three different values of $c_F$ and which all describe the inclusive jet $\Raa$ data with the same precision. To calculate the $\Raa$ in centrality bins of the published measurement \cite{ATLAS:2023iad}, we use the Glauber model to estimate the average path length for those centrality bins and the above-given parameterization of $\pars(\parl)$ dependence. The result of the calculations is shown together with the measured $\Raa$ in the upper left panel of Figure~$\ref{fig:gamma}$. One can see that the differences in $\Raa$ between different $c_F$ values are rather large, confirming that the $\gamma$-jet measurement can help constrain the role of flavor in the jet quenching. One can also see that the calculated $\Raa$ qualitatively reproduces the shape of the measured $\Raa$ in the region of $\ptjet \lesssim 120$~GeV. Namely, it reproduces the decrease of $\Raa$ at low $\ptjet$, a presence of minimum, and then a smooth rise with increasing $\pt$. None of the quenching parameterizations used in the previous section can, however, reproduce the local maximum present in the data at $\ptjet \approx 150$~GeV and the subsequent decrease. Since this local maximum is present for all the centralities, we may speculate that this increase is connected with fluctuations in the $pp$ reference \footnote{The presence of this local maximum is not significant with respect to uncertainties delivered by the experiment. Nevertheless, it is also interesting to notice that Figure~1 in Ref.~\cite{ATLAS:2023iad} contains a fluctuation in quark jet fraction in PYTHIA8 in a similar $\ptjet$ window. Such fluctuation in MC reference could, in principle, affect, e.g., the unfolding of the data.}. While the parametric modeling can describe quite successfully the shape of the measured $\Raa$, especially in central collisions, it noticeably fails in describing the overall magnitude of $\Raa$ in $0-10\%$ and $10-30\%$ centrality bins. 

To further explore the building blocks needed to reproduce the measured $\Raa$ we evaluate the $\Raa$ for the same jet quenching parameters but different input spectra of jets in the $\gamma$-jet system. We compare the following inputs: PYTHIA8 with no reweighting applied to match the MC jet spectra to those measured in the $pp$ data; PYTHIA8 with reweighting applied only on direct photons; PYTHIA8 with reweighting applied only on photons from fragmentation; PYTHIA8 with no nPDFs (all previously described versions apply nPDFs for the spectra entering the numerator of $\Raa$); and HERWIG7 with no nPDFs. A comparison of the five calculated $\Raa$ distributions is shown in the upper right panel of Figure~\ref{fig:gamma}. One can see that reweighting of jet spectra to match those measured in the data influences the magnitude of the calculated $\Raa$ by less than 10\% (and slightly more in the case of HERWIG, which we do not show). Implementation of nPDF effects influences the magnitude of $\Raa$ more substantially, by 15-20\%. The choice of MC generator (PYTHIA8 vs HERWIG7) influences the magnitude of $\Raa$ by another $\approx 10\%$. We may, therefore, conclude that the precise knowledge of input parton spectra plays an important role in the ability to determine the exact shape of $\Raa$. At the same time, the lack of full knowledge of input parton spectra cannot fully explain the difference in the magnitude of $\Raa$ among different centrality classes. 

A possible source of the apparently smaller magnitude of the energy loss suffered by jets in the $\gamma$-jet system than in the inclusive jet system may be the selection bias \cite{Renk:2012ve,STAR:2011fmy}. Both jets in the dijet system, which dominate the inclusive jet sample, suffer the energy loss, while in the case of the $\gamma$-jets, it is only one jet that interacts with the medium, whereas the $\gamma$ remains unresolved by the medium. We will not attempt to model the selection bias here. We will only quantify the change in the average path length needed to achieve a good description of the measured $\Raa$. This is done by minimizing differences between the model and the data while changing the magnitude of the quenching dictated by the parameter $s$. We can then translate this change to a change in the average path-length traveled by jets in the case of $\gamma$-jet events, $\langle L_{\gamma} \rangle$, and average path-length traveled in the case of inclusive jet events, $\parl$. The $\langle L_{\gamma} \rangle/\parl$ ratios are found to be the following: $0.80\pm0.02$, $0.90\pm0.03$, and $1.07\pm0.03$ for $0-10\%$, $10-30\%$, and $30-80\%$ centrality collisions, respectively. This $\Raa$ is then shown in the lower left panel of Figure~\ref{fig:gamma}. One can see a good model-to-data agreement for the $\Raa$ region outside of the second maximum. The above-calculated ratios may then be valuable for constraining models implementing the differences between the inclusive jet suppression and $\gamma$-jet suppression at the microscopic level.

While the change in the path-length dependence of energy loss may be an acceptable source of the unexpected centrality dependence of the energy loss in $\gamma$-jet events, it is also useful to discuss conceivable contamination of signal by the inclusive jet background. The ratio of $\gamma$-jet to inclusive jet cross-sections determined from $pp$ data \cite{ATLAS:2023iad} is $\approx 2.5\cdot10^{-4}$ for jets with $50 < \pt < 100$~GeV. This rather large factor raises a question about whether the photon selection could be contaminated by isolated, predominantly neutral hadrons that could mimic a photon. 
To quantify this contribution, we use the large sample of PYTHIA dijet events (described in Section~\ref{sec:spec}) to find the cross-section for the production of jets opposite isolated hadrons with $\pt>50$~GeV. The isolation criteria are the same as those applied previously on photons (described in Section~\ref{sec:spec}). The ratio of the cross-section for producing the jet with $50< \pt < 100$~GeV in the sample with isolated hadrons and in the inclusive jet sample is found to be $1.4 \cdot 10^{-4}$ and $5.6 \cdot 10^{-5}$ for isolated hadrons of any charge and for predominately neutral hadrons, respectively. Predominantly neutral hadrons are a mixture of neutral hadrons and charged hadrons that are simulated to be missed by tracking due to finite charged-particle reconstruction efficiency (here taken to be 80\%~\cite{ATLAS:2018bvp}). Given the uncertainty of PYTHIA to model the very end of the fragmentation spectrum and given the fact that these cross-sections are of a similar order of magnitude as those for the measured cross-section of jets in the $\gamma$-jet system, one might conclude that the contamination by inclusive jets is plausible. 
We should, however, stress that this contribution is eliminated by experiments by performing sophisticated shower-shape cuts and double-sideband subtraction, which we cannot apply in this analysis. Additionally, any residual contamination is covered by experiments in systematic uncertainties \cite{ATLAS:2017nah,ATLAS:2016ecu,CMS:2020uim}.
While this hadronic contribution is eliminated by experiments, it is still useful to quantify the jet $\Raa$ for jets in the isolated-hadron-jet system. The resulting $\Raa$ is shown in the lower right panel of Figure~\ref{fig:gamma}. In that evaluation, we fixed the values of $\alpha$ and $c_F$ coefficients to be the same as in the main study and only optimized the values of $s$ parameter by the same minimization procedure as the one used before. One can see that the shape of the resulting $\Raa$ is similar to that obtained for jets from the $\gamma$-jet system.

\section{Summary and conclusions}
\label{sec:sum}

The parametric approach to jet quenching modeling was used to study jet quenching in inclusive jet, $b$-jet, and $\gamma$-jet systems. Various parameterizations of $\ptjet$ dependence of energy loss were studied, and it was shown that 
a pure power-law ansatz for $\ptjet$-dependence of the energy loss (Equation~(\ref{eq:s})) together with a simple parameterization of energy loss fluctuations (Equation~(\ref{eq:fluct3})) provide a very good description of inclusive jet $\Raa$. Important for the agreement with the data is also the use of realistic input parton spectra with implemented isospin and nPDF effects.
The power-law index $\alpha$ in $(\ptjet)^\alpha$ dependence of energy loss was found to be centrality independent with a value of $0.27 \pm 0.03$ for color factor $c_F = 1.78$. 

Using parameterizations of inclusive jet suppression, the path-length ($L$) dependence of energy loss was quantified. Three estimates of average path length were used. One based on the Glauber model and two based on calculations with Trajectum. 
The power-law index $\delta$ in $L^\delta$ dependence of energy loss was found to be $1.55 \pm 0.13$ for the Trajectum model, which takes into account the effects of fluid motion with respect to the penetrating probe. The $\delta$ was found to be $2.01 \pm 0.17$ and $2.10 \pm 0.16$ for Glauber and Trajectum without effects of co-moving fluid. The values of $\delta$ were found to be up to 15\% higher when requiring zero energy loss at $L=0$.  
These results thus support the physics picture of the radiative nature of parton energy loss.

Using the $\ptjet$ and path-length parameterizations of the energy loss, jet $v_2$ was studied. It was demonstrated that the overall magnitude of jet $v_2$ is highly sensitive to differences between in-plane and out-of-plane path length estimates, which differ significantly among the used models and which may substantially affect the value of the extracted coefficient $\delta$.  

The $\ptjet$ and path-length parameterizations of the energy loss were then used to predict the baseline magnitude of jet $\Raa$ in 7 TeV oxygen-oxygen collisions. The $b$-jet suppression was also briefly discussed, showing statistically insignificant differences in the path-length dependence with respect to the inclusive jet suppression. 

The $\ptjet$ and path-length parameterizations of the energy loss were further used to study the jet suppression in the $\gamma$-jet system.  It was shown that the differences between the suppression of quark-initiated and gluon-initiated jets may explain the shape of the $\Raa$ of jets in the $\gamma$-jet system in all centrality bins and its magnitude in the centrality bin of $30-80\%$, but more detailed knowledge of jet quenching at the microscopic level is needed to understand the apparent lack of suppression in $0-10\%$ and $10-30\%$ central collisions. 

Since the parametric approach used in this paper represents a framework with only minimal assumptions on the underlying jet-quenching mechanism, the provided results may be used to help guide the full microscopic modeling of parton energy loss in expanding quark-gluon plasma. The presented work also demonstrates a way to further study the impact of various aspects of jet quenching physics on measured data.

\section*{Acknowledgments}
We would like to thank to Adam Takacs, Yeonju Go, Dennis Perepelitsa, Fran\c cois Arleo, Wilke van der Schee, and Xin-Nian Wang for useful discussions during the preparation of this manuscript.
This work was supported by The Ministry of Education, Youth and Sports of the Czech Republic under project ERC-CZ LL2327.

\bibliography{paper}

\begin{thebibliography}{100}
\expandafter\ifx\csname url\endcsname\relax
  \def\url#1{\texttt{#1}}\fi
\expandafter\ifx\csname urlprefix\endcsname\relax\def\urlprefix{URL }\fi
\expandafter\ifx\csname href\endcsname\relax
  \def\href#1#2{#2} \def\path#1{#1}\fi

\bibitem{Ding:2015ona}
H.-T. Ding, F.~Karsch, S.~Mukherjee, {Thermodynamics of strong-interaction
  matter from Lattice QCD}, Int. J. Mod. Phys. E 24~(10) (2015) 1530007.
\newblock \href {http://arxiv.org/abs/1504.05274} {\path{arXiv:1504.05274}},
  \href {https://doi.org/10.1142/S0218301315300076}
  {\path{doi:10.1142/S0218301315300076}}.

\bibitem{Busza:2018rrf}
W.~Busza, K.~Rajagopal, W.~van~der Schee, {Heavy Ion Collisions: The Big
  Picture, and the Big Questions}, Ann. Rev. Nucl. Part. Sci. 68 (2018)
  339--376.
\newblock \href {http://arxiv.org/abs/1802.04801} {\path{arXiv:1802.04801}},
  \href {https://doi.org/10.1146/annurev-nucl-101917-020852}
  {\path{doi:10.1146/annurev-nucl-101917-020852}}.

\bibitem{Mehtar-Tani:2013pia}
Y.~Mehtar-Tani, J.~G. Milhano, K.~Tywoniuk, {Jet physics in heavy-ion
  collisions}, Int. J. Mod. Phys. A 28 (2013) 1340013.
\newblock \href {http://arxiv.org/abs/1302.2579} {\path{arXiv:1302.2579}},
  \href {https://doi.org/10.1142/S0217751X13400137}
  {\path{doi:10.1142/S0217751X13400137}}.

\bibitem{Majumder:2010qh}
A.~Majumder, M.~Van~Leeuwen, {The Theory and Phenomenology of Perturbative QCD
  Based Jet Quenching}, Prog. Part. Nucl. Phys. 66 (2011) 41--92.
\newblock \href {http://arxiv.org/abs/1002.2206} {\path{arXiv:1002.2206}},
  \href {https://doi.org/10.1016/j.ppnp.2010.09.001}
  {\path{doi:10.1016/j.ppnp.2010.09.001}}.

\bibitem{Casalderrey-Solana:2007knd}
J.~Casalderrey-Solana, C.~A. Salgado, {Introductory lectures on jet quenching
  in heavy ion collisions}, Acta Phys. Polon. B 38 (2007) 3731--3794.
\newblock \href {http://arxiv.org/abs/0712.3443} {\path{arXiv:0712.3443}}.

\bibitem{Cao:2020wlm}
S.~Cao, X.-N. Wang, {Jet quenching and medium response in high-energy heavy-ion
  collisions: a review}, Rept. Prog. Phys. 84~(2) (2021) 024301.
\newblock \href {http://arxiv.org/abs/2002.04028} {\path{arXiv:2002.04028}},
  \href {https://doi.org/10.1088/1361-6633/abc22b}
  {\path{doi:10.1088/1361-6633/abc22b}}.

\bibitem{Apolinario:2022vzg}
L.~Apolin\'ario, Y.-J. Lee, M.~Winn, {Heavy quarks and jets as probes of the
  QGP}, Prog. Part. Nucl. Phys. 127 (2022) 103990.
\newblock \href {http://arxiv.org/abs/2203.16352} {\path{arXiv:2203.16352}},
  \href {https://doi.org/10.1016/j.ppnp.2022.103990}
  {\path{doi:10.1016/j.ppnp.2022.103990}}.

\bibitem{Cunqueiro:2021wls}
L.~Cunqueiro, A.~M. Sickles, {Studying the QGP with Jets at the LHC and RHIC},
  Prog. Part. Nucl. Phys. 124 (2022) 103940.
\newblock \href {http://arxiv.org/abs/2110.14490} {\path{arXiv:2110.14490}},
  \href {https://doi.org/10.1016/j.ppnp.2022.103940}
  {\path{doi:10.1016/j.ppnp.2022.103940}}.

\bibitem{Connors:2017ptx}
M.~Connors, C.~Nattrass, R.~Reed, S.~Salur, {Jet measurements in heavy ion
  physics}, Rev. Mod. Phys. 90 (2018) 025005.
\newblock \href {http://arxiv.org/abs/1705.01974} {\path{arXiv:1705.01974}},
  \href {https://doi.org/10.1103/RevModPhys.90.025005}
  {\path{doi:10.1103/RevModPhys.90.025005}}.

\bibitem{Citron:2018lsq}
Z.~Citron, et~al., {Report from Working Group 5}: {Future physics opportunities
  for high-density QCD at the LHC with heavy-ion and proton beams}, CERN Yellow
  Rep. Monogr. 7 (2019) 1159--1410.
\newblock \href {http://arxiv.org/abs/1812.06772} {\path{arXiv:1812.06772}},
  \href {https://doi.org/10.23731/CYRM-2019-007.1159}
  {\path{doi:10.23731/CYRM-2019-007.1159}}.

\bibitem{Dissertori2009}
G.~Dissertori, I.~G. Knowles, M.~Schmelling, Quantum Chromodynamics: High
  energy experiments and theory, Oxford University Press, 2009.
\newblock \href {https://doi.org/10.1093/acprof:oso/9780199566419.001.0001}
  {\path{doi:10.1093/acprof:oso/9780199566419.001.0001}}.

\bibitem{Gallicchio:2011xq}
J.~Gallicchio, M.~D. Schwartz, {Quark and Gluon Tagging at the LHC}, Phys. Rev.
  Lett. 107 (2011) 172001.
\newblock \href {http://arxiv.org/abs/1106.3076} {\path{arXiv:1106.3076}},
  \href {https://doi.org/10.1103/PhysRevLett.107.172001}
  {\path{doi:10.1103/PhysRevLett.107.172001}}.

\bibitem{Mehtar-Tani:2018zba}
Y.~Mehtar-Tani, S.~Schlichting, {Universal quark to gluon ratio in
  medium-induced parton cascade}, JHEP 09 (2018) 144.
\newblock \href {http://arxiv.org/abs/1807.06181} {\path{arXiv:1807.06181}},
  \href {https://doi.org/10.1007/JHEP09(2018)144}
  {\path{doi:10.1007/JHEP09(2018)144}}.

\bibitem{Chien:2015hda}
Y.-T. Chien, I.~Vitev, {Towards the understanding of jet shapes and cross
  sections in heavy ion collisions using soft-collinear effective theory}, JHEP
  05 (2016) 023.
\newblock \href {http://arxiv.org/abs/1509.07257} {\path{arXiv:1509.07257}},
  \href {https://doi.org/10.1007/JHEP05(2016)023}
  {\path{doi:10.1007/JHEP05(2016)023}}.

\bibitem{Casalderrey-Solana:2015vaa}
J.~Casalderrey-Solana, D.~C. Gulhan, J.~G. Milhano, D.~Pablos, K.~Rajagopal,
  {Predictions for Boson-Jet Observables and Fragmentation Function Ratios from
  a Hybrid Strong/Weak Coupling Model for Jet Quenching}, JHEP 03 (2016) 053.
\newblock \href {http://arxiv.org/abs/1508.00815} {\path{arXiv:1508.00815}},
  \href {https://doi.org/10.1007/JHEP03(2016)053}
  {\path{doi:10.1007/JHEP03(2016)053}}.

\bibitem{Zapp:2008gi}
K.~Zapp, G.~Ingelman, J.~Rathsman, J.~Stachel, U.~A. Wiedemann, {A Monte Carlo
  Model for 'Jet Quenching'}, Eur. Phys. J. C 60 (2009) 617--632.
\newblock \href {http://arxiv.org/abs/0804.3568} {\path{arXiv:0804.3568}},
  \href {https://doi.org/10.1140/epjc/s10052-009-0941-2}
  {\path{doi:10.1140/epjc/s10052-009-0941-2}}.

\bibitem{Armesto:2009fj}
N.~Armesto, L.~Cunqueiro, C.~A. Salgado, {Q-PYTHIA: A Medium-modified
  implementation of final state radiation}, Eur. Phys. J. C 63 (2009) 679--690.
\newblock \href {http://arxiv.org/abs/0907.1014} {\path{arXiv:0907.1014}},
  \href {https://doi.org/10.1140/epjc/s10052-009-1133-9}
  {\path{doi:10.1140/epjc/s10052-009-1133-9}}.

\bibitem{Casalderrey-Solana:2014bpa}
J.~Casalderrey-Solana, D.~C. Gulhan, J.~G. Milhano, D.~Pablos, K.~Rajagopal, {A
  Hybrid Strong/Weak Coupling Approach to Jet Quenching}, JHEP 10 (2014) 019,
  [Erratum: JHEP 09, 175 (2015)].
\newblock \href {http://arxiv.org/abs/1405.3864} {\path{arXiv:1405.3864}},
  \href {https://doi.org/10.1007/JHEP09(2015)175}
  {\path{doi:10.1007/JHEP09(2015)175}}.

\bibitem{Schenke:2009gb}
B.~Schenke, C.~Gale, S.~Jeon, {MARTINI: An Event generator for relativistic
  heavy-ion collisions}, Phys. Rev. C 80 (2009) 054913.
\newblock \href {http://arxiv.org/abs/0909.2037} {\path{arXiv:0909.2037}},
  \href {https://doi.org/10.1103/PhysRevC.80.054913}
  {\path{doi:10.1103/PhysRevC.80.054913}}.

\bibitem{Majumder:2013re}
A.~Majumder, {Incorporating Space-Time Within Medium-Modified Jet Event
  Generators}, Phys. Rev. C 88 (2013) 014909.
\newblock \href {http://arxiv.org/abs/1301.5323} {\path{arXiv:1301.5323}},
  \href {https://doi.org/10.1103/PhysRevC.88.014909}
  {\path{doi:10.1103/PhysRevC.88.014909}}.

\bibitem{Wang:2013cia}
X.-N. Wang, Y.~Zhu, {Medium Modification of $\gamma$-jets in High-energy
  Heavy-ion Collisions}, Phys. Rev. Lett. 111~(6) (2013) 062301.
\newblock \href {http://arxiv.org/abs/1302.5874} {\path{arXiv:1302.5874}},
  \href {https://doi.org/10.1103/PhysRevLett.111.062301}
  {\path{doi:10.1103/PhysRevLett.111.062301}}.

\bibitem{Mehtar-Tani:2010ebp}
Y.~Mehtar-Tani, C.~A. Salgado, K.~Tywoniuk, {Anti-angular ordering of gluon
  radiation in QCD media}, Phys. Rev. Lett. 106 (2011) 122002.
\newblock \href {http://arxiv.org/abs/1009.2965} {\path{arXiv:1009.2965}},
  \href {https://doi.org/10.1103/PhysRevLett.106.122002}
  {\path{doi:10.1103/PhysRevLett.106.122002}}.

\bibitem{Casalderrey-Solana:2012evi}
J.~Casalderrey-Solana, Y.~Mehtar-Tani, C.~A. Salgado, K.~Tywoniuk, {New picture
  of jet quenching dictated by color coherence}, Phys. Lett. B 725 (2013)
  357--360.
\newblock \href {http://arxiv.org/abs/1210.7765} {\path{arXiv:1210.7765}},
  \href {https://doi.org/10.1016/j.physletb.2013.07.046}
  {\path{doi:10.1016/j.physletb.2013.07.046}}.

\bibitem{Mehtar-Tani:2011hma}
Y.~Mehtar-Tani, C.~A. Salgado, K.~Tywoniuk, {Jets in QCD Media: From Color
  Coherence to Decoherence}, Phys. Lett. B 707 (2012) 156--159.
\newblock \href {http://arxiv.org/abs/1102.4317} {\path{arXiv:1102.4317}},
  \href {https://doi.org/10.1016/j.physletb.2011.12.042}
  {\path{doi:10.1016/j.physletb.2011.12.042}}.

\bibitem{Mehtar-Tani:2011vlz}
Y.~Mehtar-Tani, K.~Tywoniuk, {Jet coherence in QCD media: the antenna radiation
  spectrum}, JHEP 01 (2013) 031.
\newblock \href {http://arxiv.org/abs/1105.1346} {\path{arXiv:1105.1346}},
  \href {https://doi.org/10.1007/JHEP01(2013)031}
  {\path{doi:10.1007/JHEP01(2013)031}}.

\bibitem{Apolinario:2014csa}
L.~Apolin\'ario, N.~Armesto, J.~G. Milhano, C.~A. Salgado, {Medium-induced
  gluon radiation and colour decoherence beyond the soft approximation}, JHEP
  02 (2015) 119.
\newblock \href {http://arxiv.org/abs/1407.0599} {\path{arXiv:1407.0599}},
  \href {https://doi.org/10.1007/JHEP02(2015)119}
  {\path{doi:10.1007/JHEP02(2015)119}}.

\bibitem{Mehtar-Tani:2017web}
Y.~Mehtar-Tani, K.~Tywoniuk, {Sudakov suppression of jets in QCD media}, Phys.
  Rev. D 98~(5) (2018) 051501.
\newblock \href {http://arxiv.org/abs/1707.07361} {\path{arXiv:1707.07361}},
  \href {https://doi.org/10.1103/PhysRevD.98.051501}
  {\path{doi:10.1103/PhysRevD.98.051501}}.

\bibitem{Barata:2021byj}
J.~a. Barata, F.~Dom\'\i{}nguez, C.~A. Salgado, V.~Vila, {A modified in-medium
  evolution equation with color coherence}, JHEP 05 (2021) 148.
\newblock \href {http://arxiv.org/abs/2101.12135} {\path{arXiv:2101.12135}},
  \href {https://doi.org/10.1007/JHEP05(2021)148}
  {\path{doi:10.1007/JHEP05(2021)148}}.

\bibitem{Caucal:2019uvr}
P.~Caucal, E.~Iancu, G.~Soyez, {Deciphering the $z_g$ distribution in
  ultrarelativistic heavy ion collisions}, JHEP 10 (2019) 273.
\newblock \href {http://arxiv.org/abs/1907.04866} {\path{arXiv:1907.04866}},
  \href {https://doi.org/10.1007/JHEP10(2019)273}
  {\path{doi:10.1007/JHEP10(2019)273}}.

\bibitem{JETSCAPE:2022jer}
A.~Kumar, et~al., {Inclusive jet and hadron suppression in a multistage
  approach}, Phys. Rev. C 107~(3) (2023) 034911.
\newblock \href {http://arxiv.org/abs/2204.01163} {\path{arXiv:2204.01163}},
  \href {https://doi.org/10.1103/PhysRevC.107.034911}
  {\path{doi:10.1103/PhysRevC.107.034911}}.

\bibitem{Cunqueiro:2023vxl}
L.~Cunqueiro, D.~Pablos, A.~Soto-Ontoso, M.~Spousta, A.~Takacs, M.~Verweij,
  {Isolating perturbative QCD splittings in heavy-ion collisions}, Phys. Rev. D
  110~(1) (2024) 014015.
\newblock \href {http://arxiv.org/abs/2311.07643} {\path{arXiv:2311.07643}},
  \href {https://doi.org/10.1103/PhysRevD.110.014015}
  {\path{doi:10.1103/PhysRevD.110.014015}}.

\bibitem{Caucal:2018dla}
P.~Caucal, E.~Iancu, A.~H. Mueller, G.~Soyez, {Vacuum-like jet fragmentation in
  a dense QCD medium}, Phys. Rev. Lett. 120 (2018) 232001.
\newblock \href {http://arxiv.org/abs/1801.09703} {\path{arXiv:1801.09703}},
  \href {https://doi.org/10.1103/PhysRevLett.120.232001}
  {\path{doi:10.1103/PhysRevLett.120.232001}}.

\bibitem{Pablos:2022mrx}
D.~Pablos, A.~Soto-Ontoso, {Pushing forward jet substructure measurements in
  heavy-ion collisions}, Phys. Rev. D 107~(9) (2023) 094003.
\newblock \href {http://arxiv.org/abs/2210.07901} {\path{arXiv:2210.07901}},
  \href {https://doi.org/10.1103/PhysRevD.107.094003}
  {\path{doi:10.1103/PhysRevD.107.094003}}.

\bibitem{Wang:1996yh}
X.-N. Wang, Z.~Huang, I.~Sarcevic, {Jet quenching in the opposite direction of
  a tagged photon in high-energy heavy ion collisions}, Phys. Rev. Lett. 77
  (1996) 231--234.
\newblock \href {http://arxiv.org/abs/hep-ph/9605213}
  {\path{arXiv:hep-ph/9605213}}, \href
  {https://doi.org/10.1103/PhysRevLett.77.231}
  {\path{doi:10.1103/PhysRevLett.77.231}}.

\bibitem{ATLAS:2023iad}
{ATLAS Collaboration}, {Comparison of inclusive and photon-tagged jet
  suppression in 5.02 TeV Pb+Pb collisions with ATLAS}, Phys. Lett. B 846
  (2023) 138154.
\newblock \href {http://arxiv.org/abs/2303.10090} {\path{arXiv:2303.10090}},
  \href {https://doi.org/10.1016/j.physletb.2023.138154}
  {\path{doi:10.1016/j.physletb.2023.138154}}.

\bibitem{CMS:2012ytf}
{CMS Collaboration}, {Studies of Jet Quenching using Isolated-Photon + Jet
  Correlations in PbPb and $pp$ Collisions at $\sqrt{s_{NN}}=2.76$ TeV}, Phys.
  Lett. B 718 (2013) 773--794.
\newblock \href {http://arxiv.org/abs/1205.0206} {\path{arXiv:1205.0206}},
  \href {https://doi.org/10.1016/j.physletb.2012.11.003}
  {\path{doi:10.1016/j.physletb.2012.11.003}}.

\bibitem{CMS:2017ehl}
{CMS Collaboration}, {Study of jet quenching with isolated-photon+jet
  correlations in PbPb and pp collisions at $\sqrt{s_{_{\mathrm{NN}}}} =$ 5.02
  TeV}, Phys. Lett. B 785 (2018) 14--39.
\newblock \href {http://arxiv.org/abs/1711.09738} {\path{arXiv:1711.09738}},
  \href {https://doi.org/10.1016/j.physletb.2018.07.061}
  {\path{doi:10.1016/j.physletb.2018.07.061}}.

\bibitem{CMS:2018jco}
{CMS Collaboration}, {Jet Shapes of Isolated Photon-Tagged Jets in Pb-Pb and pp
  Collisions at $\sqrt{s_\mathrm{NN}} =$ 5.02 TeV}, Phys. Rev. Lett. 122~(15)
  (2019) 152001.
\newblock \href {http://arxiv.org/abs/1809.08602} {\path{arXiv:1809.08602}},
  \href {https://doi.org/10.1103/PhysRevLett.122.152001}
  {\path{doi:10.1103/PhysRevLett.122.152001}}.

\bibitem{CMS:2018mqn}
{CMS Collaboration}, {Observation of Medium-Induced Modifications of Jet
  Fragmentation in Pb-Pb Collisions at $\sqrt{s_{NN}}=$ 5.02 TeV Using Isolated
  Photon-Tagged Jets}, Phys. Rev. Lett. 121~(24) (2018) 242301.
\newblock \href {http://arxiv.org/abs/1801.04895} {\path{arXiv:1801.04895}},
  \href {https://doi.org/10.1103/PhysRevLett.121.242301}
  {\path{doi:10.1103/PhysRevLett.121.242301}}.

\bibitem{ATLAS:2018dgb}
{ATLAS Collaboration}, {Measurement of photon-jet transverse momentum
  correlations in 5.02 TeV Pb+Pb and $pp$ collisions with ATLAS}, Phys. Lett. B
  789 (2019) 167--190.
\newblock \href {http://arxiv.org/abs/1809.07280} {\path{arXiv:1809.07280}},
  \href {https://doi.org/10.1016/j.physletb.2018.12.023}
  {\path{doi:10.1016/j.physletb.2018.12.023}}.

\bibitem{ATLAS:2019dsv}
{ATLAS Collaboration}, {Comparison of Fragmentation Functions for Jets
  Dominated by Light Quarks and Gluons from $pp$ and Pb+Pb Collisions in
  ATLAS}, Phys. Rev. Lett. 123~(4) (2019) 042001.
\newblock \href {http://arxiv.org/abs/1902.10007} {\path{arXiv:1902.10007}},
  \href {https://doi.org/10.1103/PhysRevLett.123.042001}
  {\path{doi:10.1103/PhysRevLett.123.042001}}.

\bibitem{Liu:2022yjc}
{Liu, A et al., ALICE Collaboration}, {Isolated Photon-Jet Correlations in
  Pb-Pb Collisions at $\sqrt {s_{NN}} = 5.02$ TeV in ALICE}, Acta Phys. Polon.
  Supp. 16~(1) (2023) 1--A54.
\newblock \href {http://arxiv.org/abs/2208.08523} {\path{arXiv:2208.08523}},
  \href {https://doi.org/10.5506/APhysPolBSupp.16.1-A54}
  {\path{doi:10.5506/APhysPolBSupp.16.1-A54}}.

\bibitem{Zhang:2009rn}
H.~Zhang, J.~F. Owens, E.~Wang, X.-N. Wang, {Tomography of high-energy nuclear
  collisions with photon-hadron correlations}, Phys. Rev. Lett. 103 (2009)
  032302.
\newblock \href {http://arxiv.org/abs/0902.4000} {\path{arXiv:0902.4000}},
  \href {https://doi.org/10.1103/PhysRevLett.103.032302}
  {\path{doi:10.1103/PhysRevLett.103.032302}}.

\bibitem{Qin:2009bk}
G.-Y. Qin, J.~Ruppert, C.~Gale, S.~Jeon, G.~D. Moore, {Jet energy loss, photon
  production, and photon-hadron correlations at RHIC}, Phys. Rev. C 80 (2009)
  054909.
\newblock \href {http://arxiv.org/abs/0906.3280} {\path{arXiv:0906.3280}},
  \href {https://doi.org/10.1103/PhysRevC.80.054909}
  {\path{doi:10.1103/PhysRevC.80.054909}}.

\bibitem{Renk:2006qg}
T.~Renk, {Towards jet tomography: gamma-hadron correlations}, Phys. Rev. C 74
  (2006) 034906.
\newblock \href {http://arxiv.org/abs/hep-ph/0607166}
  {\path{arXiv:hep-ph/0607166}}, \href
  {https://doi.org/10.1103/PhysRevC.74.034906}
  {\path{doi:10.1103/PhysRevC.74.034906}}.

\bibitem{Betz:2014cza}
B.~Betz, M.~Gyulassy, {Constraints on the Path-Length Dependence of Jet
  Quenching in Nuclear Collisions at RHIC and LHC}, JHEP 08 (2014) 090,
  [Erratum: JHEP 10, 043 (2014)].
\newblock \href {http://arxiv.org/abs/1404.6378} {\path{arXiv:1404.6378}},
  \href {https://doi.org/10.1007/JHEP10(2014)043}
  {\path{doi:10.1007/JHEP10(2014)043}}.

\bibitem{Betz:2016ayq}
B.~Betz, M.~Gyulassy, M.~Luzum, J.~Noronha, J.~Noronha-Hostler, I.~Portillo,
  C.~Ratti, {Cumulants and nonlinear response of high $p_T$ harmonic flow at
  $\sqrt{s_{NN}}=5.02$ TeV}, Phys. Rev. C 95~(4) (2017) 044901.
\newblock \href {http://arxiv.org/abs/1609.05171} {\path{arXiv:1609.05171}},
  \href {https://doi.org/10.1103/PhysRevC.95.044901}
  {\path{doi:10.1103/PhysRevC.95.044901}}.

\bibitem{Djordjevic:2018ita}
M.~Djordjevic, D.~Zigic, M.~Djordjevic, J.~Auvinen, {How to test path-length
  dependence in energy loss mechanisms: analysis leading to a new observable},
  Phys. Rev. C 99~(6) (2019) 061902.
\newblock \href {http://arxiv.org/abs/1805.04030} {\path{arXiv:1805.04030}},
  \href {https://doi.org/10.1103/PhysRevC.99.061902}
  {\path{doi:10.1103/PhysRevC.99.061902}}.

\bibitem{Arleo:2022shs}
F.~Arleo, G.~Falmagne, {Probing the path-length dependence of parton energy
  loss via scaling properties in heavy ion collisions}, Phys. Rev. D 109~(5)
  (2024) L051503.
\newblock \href {http://arxiv.org/abs/2212.01324} {\path{arXiv:2212.01324}},
  \href {https://doi.org/10.1103/PhysRevD.109.L051503}
  {\path{doi:10.1103/PhysRevD.109.L051503}}.

\bibitem{Spousta:2015fca}
M.~Spousta, B.~Cole, {Interpreting single jet measurements in Pb $+$ Pb
  collisions at the LHC}, Eur. Phys. J. C 76~(2) (2016) 50.
\newblock \href {http://arxiv.org/abs/1504.05169} {\path{arXiv:1504.05169}},
  \href {https://doi.org/10.1140/epjc/s10052-016-3896-0}
  {\path{doi:10.1140/epjc/s10052-016-3896-0}}.

\bibitem{Spousta:2016agr}
M.~Spousta, {On similarity of jet quenching and charmonia suppression}, Phys.
  Lett. B 767 (2017) 10--15.
\newblock \href {http://arxiv.org/abs/1606.00903} {\path{arXiv:1606.00903}},
  \href {https://doi.org/10.1016/j.physletb.2017.01.041}
  {\path{doi:10.1016/j.physletb.2017.01.041}}.

\bibitem{Takacs:2021bpv}
A.~Takacs, K.~Tywoniuk, {Quenching effects in the cumulative jet spectrum},
  JHEP 10 (2021) 038.
\newblock \href {http://arxiv.org/abs/2103.14676} {\path{arXiv:2103.14676}},
  \href {https://doi.org/10.1007/JHEP10(2021)038}
  {\path{doi:10.1007/JHEP10(2021)038}}.

\bibitem{Adhya:2021kws}
S.~P. Adhya, C.~A. Salgado, M.~Spousta, K.~Tywoniuk, {Multi-partonic medium
  induced cascades in expanding media}, Eur. Phys. J. C 82~(1) (2022) 20.
\newblock \href {http://arxiv.org/abs/2106.02592} {\path{arXiv:2106.02592}},
  \href {https://doi.org/10.1140/epjc/s10052-021-09950-8}
  {\path{doi:10.1140/epjc/s10052-021-09950-8}}.

\bibitem{Mehtar-Tani:2022zwf}
Y.~Mehtar-Tani, S.~Schlichting, I.~Soudi, {Jet thermalization in QCD kinetic
  theory}, JHEP 05 (2023) 091.
\newblock \href {http://arxiv.org/abs/2209.10569} {\path{arXiv:2209.10569}},
  \href {https://doi.org/10.1007/JHEP05(2023)091}
  {\path{doi:10.1007/JHEP05(2023)091}}.

\bibitem{Caucal:2020uic}
P.~Caucal, E.~Iancu, G.~Soyez, {Jet radiation in a longitudinally expanding
  medium}, JHEP 04 (2021) 209.
\newblock \href {http://arxiv.org/abs/2012.01457} {\path{arXiv:2012.01457}},
  \href {https://doi.org/10.1007/JHEP04(2021)209}
  {\path{doi:10.1007/JHEP04(2021)209}}.

\bibitem{Ke:2020clc}
W.~Ke, X.-N. Wang, {QGP modification to single inclusive jets in a calibrated
  transport model}, JHEP 05 (2021) 041.
\newblock \href {http://arxiv.org/abs/2010.13680} {\path{arXiv:2010.13680}},
  \href {https://doi.org/10.1007/JHEP05(2021)041}
  {\path{doi:10.1007/JHEP05(2021)041}}.

\bibitem{Qiu:2019sfj}
J.-W. Qiu, F.~Ringer, N.~Sato, P.~Zurita, {Factorization of jet cross sections
  in heavy-ion collisions}, Phys. Rev. Lett. 122~(25) (2019) 252301.
\newblock \href {http://arxiv.org/abs/1903.01993} {\path{arXiv:1903.01993}},
  \href {https://doi.org/10.1103/PhysRevLett.122.252301}
  {\path{doi:10.1103/PhysRevLett.122.252301}}.

\bibitem{He:2018gks}
Y.~He, L.-G. Pang, X.-N. Wang, {Bayesian extraction of jet energy loss
  distributions in heavy-ion collisions}, Phys. Rev. Lett. 122~(25) (2019)
  252302.
\newblock \href {http://arxiv.org/abs/1808.05310} {\path{arXiv:1808.05310}},
  \href {https://doi.org/10.1103/PhysRevLett.122.252302}
  {\path{doi:10.1103/PhysRevLett.122.252302}}.

\bibitem{Zhang:2023oid}
S.-L. Zhang, E.~Wang, H.~Xing, B.-W. Zhang, {Flavor dependence of jet quenching
  in heavy-ion collisions from a Bayesian analysis}, Phys. Lett. B 850 (2024)
  138549.
\newblock \href {http://arxiv.org/abs/2303.14881} {\path{arXiv:2303.14881}},
  \href {https://doi.org/10.1016/j.physletb.2024.138549}
  {\path{doi:10.1016/j.physletb.2024.138549}}.

\bibitem{He:2018xjv}
Y.~He, S.~Cao, W.~Chen, T.~Luo, L.-G. Pang, X.-N. Wang, {Interplaying
  mechanisms behind single inclusive jet suppression in heavy-ion collisions},
  Phys. Rev. C 99~(5) (2019) 054911.
\newblock \href {http://arxiv.org/abs/1809.02525} {\path{arXiv:1809.02525}},
  \href {https://doi.org/10.1103/PhysRevC.99.054911}
  {\path{doi:10.1103/PhysRevC.99.054911}}.

\bibitem{Wu:2023azi}
J.~Wu, W.~Ke, X.-N. Wang, {Bayesian inference of the path-length dependence of
  jet energy loss}, Phys. Rev. C 108~(3) (2023) 034911.
\newblock \href {http://arxiv.org/abs/2304.06339} {\path{arXiv:2304.06339}},
  \href {https://doi.org/10.1103/PhysRevC.108.034911}
  {\path{doi:10.1103/PhysRevC.108.034911}}.

\bibitem{Brewer:2021hmh}
J.~Brewer, Q.~Brodsky, K.~Rajagopal, {Disentangling jet modification in jet
  simulations and in Z+jet data}, JHEP 02 (2022) 175.
\newblock \href {http://arxiv.org/abs/2110.13159} {\path{arXiv:2110.13159}},
  \href {https://doi.org/10.1007/JHEP02(2022)175}
  {\path{doi:10.1007/JHEP02(2022)175}}.

\bibitem{Sjostrand:2014zea}
T.~Sj\"ostrand, S.~Ask, J.~R. Christiansen, R.~Corke, N.~Desai, P.~Ilten,
  S.~Mrenna, S.~Prestel, C.~O. Rasmussen, P.~Z. Skands, {An introduction to
  PYTHIA 8.2}, Comput. Phys. Commun. 191 (2015) 159--177.
\newblock \href {http://arxiv.org/abs/1410.3012} {\path{arXiv:1410.3012}},
  \href {https://doi.org/10.1016/j.cpc.2015.01.024}
  {\path{doi:10.1016/j.cpc.2015.01.024}}.

\bibitem{Bierlich:2022pfr}
C.~Bierlich, et~al., {A comprehensive guide to the physics and usage of PYTHIA
  8.3} (3 2022).
\newblock \href {http://arxiv.org/abs/2203.11601} {\path{arXiv:2203.11601}},
  \href {https://doi.org/10.21468/SciPostPhysCodeb.8}
  {\path{doi:10.21468/SciPostPhysCodeb.8}}.

\bibitem{TheATLAScollaboration:2014rfk}
{ATLAS Pythia 8 tunes to 7 TeV data} (11 2014).

\bibitem{Ball:2012cx}
R.~D. Ball, et~al., {Parton distributions with LHC data}, Nucl. Phys. B 867
  (2013) 244--289.
\newblock \href {http://arxiv.org/abs/1207.1303} {\path{arXiv:1207.1303}},
  \href {https://doi.org/10.1016/j.nuclphysb.2012.10.003}
  {\path{doi:10.1016/j.nuclphysb.2012.10.003}}.

\bibitem{ATLAS:2023hso}
{ATLAS Collaboration}, {Measurement of Suppression of Large-Radius Jets and Its
  Dependence on Substructure in Pb+Pb Collisions at $\sqrt{s_{NN}}=5.02$~TeV
  with the ATLAS Detector}, Phys. Rev. Lett. 131~(17) (2023) 172301.
\newblock \href {http://arxiv.org/abs/2301.05606} {\path{arXiv:2301.05606}},
  \href {https://doi.org/10.1103/PhysRevLett.131.172301}
  {\path{doi:10.1103/PhysRevLett.131.172301}}.

\bibitem{Bahr:2008pv}
M.~Bahr, et~al., {Herwig++ Physics and Manual}, Eur. Phys. J. C 58 (2008)
  639--707.
\newblock \href {http://arxiv.org/abs/0803.0883} {\path{arXiv:0803.0883}},
  \href {https://doi.org/10.1140/epjc/s10052-008-0798-9}
  {\path{doi:10.1140/epjc/s10052-008-0798-9}}.

\bibitem{ATLAS:2020cli}
{ATLAS Collaboration}, {Jet energy scale and resolution measured in
  proton\textendash{}proton collisions at $\sqrt{s}=13$~TeV with the ATLAS
  detector}, Eur. Phys. J. C 81~(8) (2021) 689.
\newblock \href {http://arxiv.org/abs/2007.02645} {\path{arXiv:2007.02645}},
  \href {https://doi.org/10.1140/epjc/s10052-021-09402-3}
  {\path{doi:10.1140/epjc/s10052-021-09402-3}}.

\bibitem{CMS:2016lmd}
{CMS Collaboration}, {Jet energy scale and resolution in the CMS experiment in
  pp collisions at 8 TeV}, JINST 12~(02) (2017) P02014.
\newblock \href {http://arxiv.org/abs/1607.03663} {\path{arXiv:1607.03663}},
  \href {https://doi.org/10.1088/1748-0221/12/02/P02014}
  {\path{doi:10.1088/1748-0221/12/02/P02014}}.

\bibitem{Cacciari:2008gp}
M.~Cacciari, G.~P. Salam, G.~Soyez, {The anti-$k_t$ jet clustering algorithm},
  JHEP 04 (2008) 063.
\newblock \href {http://arxiv.org/abs/0802.1189} {\path{arXiv:0802.1189}},
  \href {https://doi.org/10.1088/1126-6708/2008/04/063}
  {\path{doi:10.1088/1126-6708/2008/04/063}}.

\bibitem{Cacciari:2011ma}
M.~Cacciari, G.~P. Salam, G.~Soyez, {FastJet User Manual}, Eur. Phys. J. C 72
  (2012) 1896.
\newblock \href {http://arxiv.org/abs/1111.6097} {\path{arXiv:1111.6097}},
  \href {https://doi.org/10.1140/epjc/s10052-012-1896-2}
  {\path{doi:10.1140/epjc/s10052-012-1896-2}}.

\bibitem{Harland_Lang_2015}
L.~A. Harland-Lang, A.~D. Martin, P.~Motylinski, R.~S. Thorne,
  \href{http://dx.doi.org/10.1140/epjc/s10052-015-3397-6}{Parton distributions
  in the lhc era: Mmht 2014 pdfs}, The European Physical Journal C 75~(5) (May
  2015).
\newblock \href {https://doi.org/10.1140/epjc/s10052-015-3397-6}
  {\path{doi:10.1140/epjc/s10052-015-3397-6}}.
\newline\urlprefix\url{http://dx.doi.org/10.1140/epjc/s10052-015-3397-6}

\bibitem{ATLAS:2018gwx}
{ATLAS Collaboration}, {Measurement of the nuclear modification factor for
  inclusive jets in Pb+Pb collisions at $\sqrt{s_\mathrm{NN}}=5.02$ TeV with
  the ATLAS detector}, Phys. Lett. B 790 (2019) 108--128.
\newblock \href {http://arxiv.org/abs/1805.05635} {\path{arXiv:1805.05635}},
  \href {https://doi.org/10.1016/j.physletb.2018.10.076}
  {\path{doi:10.1016/j.physletb.2018.10.076}}.

\bibitem{Eskola:2016oht}
K.~J. Eskola, P.~Paakkinen, H.~Paukkunen, C.~A. Salgado, {EPPS16: Nuclear
  parton distributions with LHC data}, Eur. Phys. J. C 77~(3) (2017) 163.
\newblock \href {http://arxiv.org/abs/1612.05741} {\path{arXiv:1612.05741}},
  \href {https://doi.org/10.1140/epjc/s10052-017-4725-9}
  {\path{doi:10.1140/epjc/s10052-017-4725-9}}.

\bibitem{Eskola:2021nhw}
K.~J. Eskola, P.~Paakkinen, H.~Paukkunen, C.~A. Salgado, {EPPS21: a global QCD
  analysis of nuclear PDFs}, Eur. Phys. J. C 82~(5) (2022) 413.
\newblock \href {http://arxiv.org/abs/2112.12462} {\path{arXiv:2112.12462}},
  \href {https://doi.org/10.1140/epjc/s10052-022-10359-0}
  {\path{doi:10.1140/epjc/s10052-022-10359-0}}.

\bibitem{AbdulKhalek:2022fyi}
R.~Abdul~Khalek, R.~Gauld, T.~Giani, E.~R. Nocera, T.~R. Rabemananjara,
  J.~Rojo, {nNNPDF3.0: evidence for a modified partonic structure in heavy
  nuclei}, Eur. Phys. J. C 82~(6) (2022) 507.
\newblock \href {http://arxiv.org/abs/2201.12363} {\path{arXiv:2201.12363}},
  \href {https://doi.org/10.1140/epjc/s10052-022-10417-7}
  {\path{doi:10.1140/epjc/s10052-022-10417-7}}.

\bibitem{James:1975dr}
F.~James, M.~Roos, {Minuit: A System for Function Minimization and Analysis of
  the Parameter Errors and Correlations}, Comput. Phys. Commun. 10 (1975)
  343--367.
\newblock \href {https://doi.org/10.1016/0010-4655(75)90039-9}
  {\path{doi:10.1016/0010-4655(75)90039-9}}.

\bibitem{Capella:1999ms}
A.~Capella, I.~M. Dremin, J.~W. Gary, V.~A. Nechitailo, J.~Tran Thanh~Van,
  {Evolution of average multiplicities of quark and gluon jets}, Phys. Rev. D61
  (2000) 074009.
\newblock \href {http://arxiv.org/abs/hep-ph/9910226}
  {\path{arXiv:hep-ph/9910226}}, \href
  {https://doi.org/10.1103/PhysRevD.61.074009}
  {\path{doi:10.1103/PhysRevD.61.074009}}.

\bibitem{Acosta:2004js}
D.~Acosta, et~al., {Measurement of charged particle multiplicities in gluon and
  quark jets in $p\bar{p}$ collisions at $\sqrt{s} = 1.8$ TeV}, Phys. Rev.
  Lett. 94 (2005) 171802.
\newblock \href {https://doi.org/10.1103/PhysRevLett.94.171802}
  {\path{doi:10.1103/PhysRevLett.94.171802}}.

\bibitem{Apolinario:2020nyw}
L.~Apolin\'ario, J.~a. Barata, G.~Milhano, {On the breaking of Casimir scaling
  in jet quenching}, Eur. Phys. J. C 80~(6) (2020) 586.
\newblock \href {http://arxiv.org/abs/2003.02893} {\path{arXiv:2003.02893}},
  \href {https://doi.org/10.1140/epjc/s10052-020-8133-1}
  {\path{doi:10.1140/epjc/s10052-020-8133-1}}.

\bibitem{Pablos:2019ngg}
D.~Pablos, {Jet Suppression From a Small to Intermediate to Large Radius},
  Phys. Rev. Lett. 124~(5) (2020) 052301.
\newblock \href {http://arxiv.org/abs/1907.12301} {\path{arXiv:1907.12301}},
  \href {https://doi.org/10.1103/PhysRevLett.124.052301}
  {\path{doi:10.1103/PhysRevLett.124.052301}}.

\bibitem{Baier:1996kr}
R.~Baier, Y.~L. Dokshitzer, A.~H. Mueller, S.~Peigne, D.~Schiff, {Radiative
  energy loss of high-energy quarks and gluons in a finite volume quark - gluon
  plasma}, Nucl. Phys. B 483 (1997) 291--320.
\newblock \href {http://arxiv.org/abs/hep-ph/9607355}
  {\path{arXiv:hep-ph/9607355}}, \href
  {https://doi.org/10.1016/S0550-3213(96)00553-6}
  {\path{doi:10.1016/S0550-3213(96)00553-6}}.

\bibitem{Gubser:2008as}
S.~S. Gubser, D.~R. Gulotta, S.~S. Pufu, F.~D. Rocha, {Gluon energy loss in the
  gauge-string duality}, JHEP 10 (2008) 052.
\newblock \href {http://arxiv.org/abs/0803.1470} {\path{arXiv:0803.1470}},
  \href {https://doi.org/10.1088/1126-6708/2008/10/052}
  {\path{doi:10.1088/1126-6708/2008/10/052}}.

\bibitem{Loizides:2017ack}
C.~Loizides, J.~Kamin, D.~d'Enterria, {Improved Monte Carlo Glauber predictions
  at present and future nuclear colliders}, Phys. Rev. C 97~(5) (2018) 054910,
  [Erratum: Phys.Rev.C 99, 019901 (2019)].
\newblock \href {http://arxiv.org/abs/1710.07098} {\path{arXiv:1710.07098}},
  \href {https://doi.org/10.1103/PhysRevC.97.054910}
  {\path{doi:10.1103/PhysRevC.97.054910}}.

\bibitem{ATLAS:2015rlt}
{ATLAS Collaboration}, {Centrality, rapidity and transverse momentum dependence
  of isolated prompt photon production in lead-lead collisions at
  $\sqrt{s_{\mathrm{NN}}} = 2.76$ TeV measured with the ATLAS detector}, Phys.
  Rev. C 93~(3) (2016) 034914.
\newblock \href {http://arxiv.org/abs/1506.08552} {\path{arXiv:1506.08552}},
  \href {https://doi.org/10.1103/PhysRevC.93.034914}
  {\path{doi:10.1103/PhysRevC.93.034914}}.

\bibitem{CMS:2020oen}
{CMS Collaboration}, {The production of isolated photons in PbPb and pp
  collisions at $\sqrt{s_\mathrm{NN}} =$ 5.02 TeV}, JHEP 07 (2020) 116.
\newblock \href {http://arxiv.org/abs/2003.12797} {\path{arXiv:2003.12797}},
  \href {https://doi.org/10.1007/JHEP07(2020)116}
  {\path{doi:10.1007/JHEP07(2020)116}}.

\bibitem{ATLAS:2019ibd}
{ATLAS Collaboration}, {Measurement of $W^\pm $ boson production in Pb+Pb
  collisions at $\sqrt{s_{\mathrm{NN}}} = 5.02~\text {Te}\text {V}$ with the
  ATLAS detector}, Eur. Phys. J. C 79~(11) (2019) 935.
\newblock \href {http://arxiv.org/abs/1907.10414} {\path{arXiv:1907.10414}},
  \href {https://doi.org/10.1140/epjc/s10052-019-7439-3}
  {\path{doi:10.1140/epjc/s10052-019-7439-3}}.

\bibitem{CMS:2012fgk}
{CMS Collaboration}, {Study of $W$ Boson Production in PbPb and $pp$ Collisions
  at $\sqrt{s_{NN}}=2.76$ TeV}, Phys. Lett. B 715 (2012) 66--87.
\newblock \href {http://arxiv.org/abs/1205.6334} {\path{arXiv:1205.6334}},
  \href {https://doi.org/10.1016/j.physletb.2012.07.025}
  {\path{doi:10.1016/j.physletb.2012.07.025}}.

\bibitem{ATLAS:2019maq}
{ATLAS Collaboration}, {$Z$ boson production in Pb+Pb collisions at
  $\sqrt{s_{\textrm{NN}}}$= 5.02 TeV measured by the ATLAS experiment}, Phys.
  Lett. B 802 (2020) 135262.
\newblock \href {http://arxiv.org/abs/1910.13396} {\path{arXiv:1910.13396}},
  \href {https://doi.org/10.1016/j.physletb.2020.135262}
  {\path{doi:10.1016/j.physletb.2020.135262}}.

\bibitem{CMS:2014dyj}
{CMS Collaboration}, {Study of Z production in PbPb and pp collisions at $
  \sqrt{s_{\mathrm{NN}}}=2.76 $ TeV in the dimuon and dielectron decay
  channels}, JHEP 03 (2015) 022.
\newblock \href {http://arxiv.org/abs/1410.4825} {\path{arXiv:1410.4825}},
  \href {https://doi.org/10.1007/JHEP03(2015)022}
  {\path{doi:10.1007/JHEP03(2015)022}}.

\bibitem{Nijs:2020ors}
G.~Nijs, W.~van~der Schee, U.~G\"ursoy, R.~Snellings, {Transverse Momentum
  Differential Global Analysis of Heavy-Ion Collisions}, Phys. Rev. Lett.
  126~(20) (2021) 202301.
\newblock \href {http://arxiv.org/abs/2010.15130} {\path{arXiv:2010.15130}},
  \href {https://doi.org/10.1103/PhysRevLett.126.202301}
  {\path{doi:10.1103/PhysRevLett.126.202301}}.

\bibitem{Nijs:2020roc}
G.~Nijs, W.~van~der Schee, U.~G\"ursoy, R.~Snellings, {Bayesian analysis of
  heavy ion collisions with the heavy ion computational framework Trajectum},
  Phys. Rev. C 103~(5) (2021) 054909.
\newblock \href {http://arxiv.org/abs/2010.15134} {\path{arXiv:2010.15134}},
  \href {https://doi.org/10.1103/PhysRevC.103.054909}
  {\path{doi:10.1103/PhysRevC.103.054909}}.

\bibitem{Beattie:2022ojg}
C.~Beattie, G.~Nijs, M.~Sas, W.~van~der Schee, {Hard probe path lengths and
  event-shape engineering of the quark-gluon plasma}, Phys. Lett. B 836 (2023)
  137596.
\newblock \href {http://arxiv.org/abs/2203.13265} {\path{arXiv:2203.13265}},
  \href {https://doi.org/10.1016/j.physletb.2022.137596}
  {\path{doi:10.1016/j.physletb.2022.137596}}.

\bibitem{Moreland:2014oya}
J.~S. Moreland, J.~E. Bernhard, S.~A. Bass, {Alternative ansatz to wounded
  nucleon and binary collision scaling in high-energy nuclear collisions},
  Phys. Rev. C 92~(1) (2015) 011901.
\newblock \href {http://arxiv.org/abs/1412.4708} {\path{arXiv:1412.4708}},
  \href {https://doi.org/10.1103/PhysRevC.92.011901}
  {\path{doi:10.1103/PhysRevC.92.011901}}.

\bibitem{Andres:2019eus}
C.~Andres, N.~Armesto, H.~Niemi, R.~Paatelainen, C.~A. Salgado, {Jet quenching
  as a probe of the initial stages in heavy-ion collisions}, Phys. Lett. B 803
  (2020) 135318.
\newblock \href {http://arxiv.org/abs/1902.03231} {\path{arXiv:1902.03231}},
  \href {https://doi.org/10.1016/j.physletb.2020.135318}
  {\path{doi:10.1016/j.physletb.2020.135318}}.

\bibitem{Adhya:2019qse}
S.~P. Adhya, C.~A. Salgado, M.~Spousta, K.~Tywoniuk, {Medium-induced cascade in
  expanding media}, JHEP 07 (2020) 150.
\newblock \href {http://arxiv.org/abs/1911.12193} {\path{arXiv:1911.12193}},
  \href {https://doi.org/10.1007/JHEP07(2020)150}
  {\path{doi:10.1007/JHEP07(2020)150}}.

\bibitem{Zigic:2018smz}
D.~Zigic, I.~Salom, J.~Auvinen, M.~Djordjevic, M.~Djordjevic, {DREENA-C
  framework: joint $R_{AA}$ and $v_2$ predictions and implications to QGP
  tomography}, J. Phys. G 46~(8) (2019) 085101.
\newblock \href {http://arxiv.org/abs/1805.03494} {\path{arXiv:1805.03494}},
  \href {https://doi.org/10.1088/1361-6471/ab2356}
  {\path{doi:10.1088/1361-6471/ab2356}}.

\bibitem{ATLAS:2021ktw}
{Measurements of azimuthal anisotropies of jet production in Pb+Pb collisions
  at $\sqrt{s_{NN}} =$ 5.02 TeV with the ATLAS detector}, Phys. Rev. C 105~(6)
  (2022) 064903.
\newblock \href {http://arxiv.org/abs/2111.06606} {\path{arXiv:2111.06606}},
  \href {https://doi.org/10.1103/PhysRevC.105.064903}
  {\path{doi:10.1103/PhysRevC.105.064903}}.

\bibitem{Behera:2021zhi}
D.~Behera, N.~Mallick, S.~Tripathy, S.~Prasad, A.~N. Mishra, R.~Sahoo,
  {Predictions on global properties in O+O collisions at the Large Hadron
  Collider using a multi-phase transport model}, Eur. Phys. J. A 58~(9) (2022)
  175.
\newblock \href {http://arxiv.org/abs/2110.04016} {\path{arXiv:2110.04016}},
  \href {https://doi.org/10.1140/epja/s10050-022-00823-6}
  {\path{doi:10.1140/epja/s10050-022-00823-6}}.

\bibitem{Gebhard:2024flv}
J.~Gebhard, A.~Mazeliauskas, A.~Takacs, {No-quenching baseline for energy loss
  signals in oxygen-oxygen collisions}, JHEP 04 (2025) 034.
\newblock \href {http://arxiv.org/abs/2410.22405} {\path{arXiv:2410.22405}},
  \href {https://doi.org/10.1007/JHEP04(2025)034}
  {\path{doi:10.1007/JHEP04(2025)034}}.

\bibitem{Huss:2020whe}
A.~Huss, A.~Kurkela, A.~Mazeliauskas, R.~Paatelainen, W.~van~der Schee, U.~A.
  Wiedemann, {Predicting parton energy loss in small collision systems}, Phys.
  Rev. C 103~(5) (2021) 054903.
\newblock \href {http://arxiv.org/abs/2007.13758} {\path{arXiv:2007.13758}},
  \href {https://doi.org/10.1103/PhysRevC.103.054903}
  {\path{doi:10.1103/PhysRevC.103.054903}}.

\bibitem{Behera:2023oxe}
D.~Behera, S.~Deb, C.~R. Singh, R.~Sahoo, {Characterizing nuclear modification
  effects in high-energy O-O collisions at energies available at the CERN Large
  Hadron Collider: A transport model perspective}, Phys. Rev. C 109~(1) (2024)
  014902.
\newblock \href {http://arxiv.org/abs/2308.06078} {\path{arXiv:2308.06078}},
  \href {https://doi.org/10.1103/PhysRevC.109.014902}
  {\path{doi:10.1103/PhysRevC.109.014902}}.

\bibitem{ATLAS:2022agz}
{ATLAS Collaboration}, {Measurement of the nuclear modification factor of
  $b$-jets in 5.02~TeV Pb+Pb collisions with the ATLAS detector}, Eur. Phys. J.
  C 83~(5) (2023) 438.
\newblock \href {http://arxiv.org/abs/2204.13530} {\path{arXiv:2204.13530}},
  \href {https://doi.org/10.1140/epjc/s10052-023-11427-9}
  {\path{doi:10.1140/epjc/s10052-023-11427-9}}.

\bibitem{Zigic:2018ovr}
D.~Zigic, I.~Salom, J.~Auvinen, M.~Djordjevic, M.~Djordjevic, {DREENA-B
  framework: first predictions of $R_{AA}$ and $v_2$ within dynamical energy
  loss formalism in evolving QCD medium}, Phys. Lett. B 791 (2019) 236--241.
\newblock \href {http://arxiv.org/abs/1805.04786} {\path{arXiv:1805.04786}},
  \href {https://doi.org/10.1016/j.physletb.2019.02.020}
  {\path{doi:10.1016/j.physletb.2019.02.020}}.

\bibitem{Dokshitzer:2001zm}
Y.~L. Dokshitzer, D.~E. Kharzeev, {Heavy quark colorimetry of QCD matter},
  Phys. Lett. B 519 (2001) 199--206.
\newblock \href {http://arxiv.org/abs/hep-ph/0106202}
  {\path{arXiv:hep-ph/0106202}}, \href
  {https://doi.org/10.1016/S0370-2693(01)01130-3}
  {\path{doi:10.1016/S0370-2693(01)01130-3}}.

\bibitem{Wicks:2005gt}
S.~Wicks, W.~Horowitz, M.~Djordjevic, M.~Gyulassy, {Elastic, inelastic, and
  path length fluctuations in jet tomography}, Nucl. Phys. A 784 (2007)
  426--442.
\newblock \href {http://arxiv.org/abs/nucl-th/0512076}
  {\path{arXiv:nucl-th/0512076}}, \href
  {https://doi.org/10.1016/j.nuclphysa.2006.12.048}
  {\path{doi:10.1016/j.nuclphysa.2006.12.048}}.

\bibitem{Ke:2018tsh}
W.~Ke, Y.~Xu, S.~A. Bass, {Linearized Boltzmann-Langevin model for heavy quark
  transport in hot and dense QCD matter}, Phys. Rev. C 98~(6) (2018) 064901.
\newblock \href {http://arxiv.org/abs/1806.08848} {\path{arXiv:1806.08848}},
  \href {https://doi.org/10.1103/PhysRevC.98.064901}
  {\path{doi:10.1103/PhysRevC.98.064901}}.

\bibitem{Renk:2012ve}
T.~Renk, {Biased showers: A common conceptual framework for the interpretation
  of high-$P_T$ observables in heavy-ion collisions}, Phys. Rev. C 88~(5)
  (2013) 054902.
\newblock \href {http://arxiv.org/abs/1212.0646} {\path{arXiv:1212.0646}},
  \href {https://doi.org/10.1103/PhysRevC.88.054902}
  {\path{doi:10.1103/PhysRevC.88.054902}}.

\bibitem{STAR:2011fmy}
H.~Agakishiev, et~al., {Studies of di-jet survival and surface emission bias in
  Au+Au collisions via angular correlations with respect to back-to-back
  leading hadrons}, Phys. Rev. C 83 (2011) 061901.
\newblock \href {http://arxiv.org/abs/1102.2669} {\path{arXiv:1102.2669}},
  \href {https://doi.org/10.1103/PhysRevC.83.061901}
  {\path{doi:10.1103/PhysRevC.83.061901}}.

\bibitem{ATLAS:2018bvp}
{ATLAS Collaboration}, {Measurement of jet fragmentation in Pb+Pb and $pp$
  collisions at $\sqrt{s_{NN}} = 5.02$ TeV with the ATLAS detector}, Phys. Rev.
  C 98~(2) (2018) 024908.
\newblock \href {http://arxiv.org/abs/1805.05424} {\path{arXiv:1805.05424}},
  \href {https://doi.org/10.1103/PhysRevC.98.024908}
  {\path{doi:10.1103/PhysRevC.98.024908}}.

\bibitem{ATLAS:2017nah}
{ATLAS Collaboration}, {Measurement of the cross section for inclusive
  isolated-photon production in $pp$ collisions at $\sqrt s=13$ TeV using the
  ATLAS detector}, Phys. Lett. B 770 (2017) 473--493.
\newblock \href {http://arxiv.org/abs/1701.06882} {\path{arXiv:1701.06882}},
  \href {https://doi.org/10.1016/j.physletb.2017.04.072}
  {\path{doi:10.1016/j.physletb.2017.04.072}}.

\bibitem{ATLAS:2016ecu}
{ATLAS Collaboration}, {Measurement of the photon identification efficiencies
  with the ATLAS detector using LHC Run-1 data}, Eur. Phys. J. C 76~(12) (2016)
  666.
\newblock \href {http://arxiv.org/abs/1606.01813} {\path{arXiv:1606.01813}},
  \href {https://doi.org/10.1140/epjc/s10052-016-4507-9}
  {\path{doi:10.1140/epjc/s10052-016-4507-9}}.

\bibitem{CMS:2020uim}
{CMS Collaboration}, {Electron and photon reconstruction and identification
  with the CMS experiment at the CERN LHC}, JINST 16~(05) (2021) P05014.
\newblock \href {http://arxiv.org/abs/2012.06888} {\path{arXiv:2012.06888}},
  \href {https://doi.org/10.1088/1748-0221/16/05/P05014}
  {\path{doi:10.1088/1748-0221/16/05/P05014}}.

\bibitem{Alioli:2010xa}
S.~Alioli, K.~Hamilton, P.~Nason, C.~Oleari, E.~Re, {Jet pair production in
  POWHEG}, JHEP 04 (2011) 081.
\newblock \href {http://arxiv.org/abs/1012.3380} {\path{arXiv:1012.3380}},
  \href {https://doi.org/10.1007/JHEP04(2011)081}
  {\path{doi:10.1007/JHEP04(2011)081}}.

\bibitem{ATLAS:2017bje}
{ATLAS Collaboration}, {Jet energy scale measurements and their systematic
  uncertainties in proton-proton collisions at $\sqrt{s} = 13$ TeV with the
  ATLAS detector}, Phys. Rev. D 96~(7) (2017) 072002.
\newblock \href {http://arxiv.org/abs/1703.09665} {\path{arXiv:1703.09665}},
  \href {https://doi.org/10.1103/PhysRevD.96.072002}
  {\path{doi:10.1103/PhysRevD.96.072002}}.

\end{thebibliography}
\bibliographystyle{elsarticle-num}

\appendix
\section{Jet \pt\ spectra}
\label{sec:appA}

In this Appendix, we provide a full list of parameters from parameterizations of jet \pt\ spectra discussed in Section~\ref{sec:spec}. Tables discussed below provide parameters $\beta_i$ and quark fractions $f_{q,0}$, defined in Eqs.~(\ref{eq:unmod}) and (\ref{eq:exp}). Besides those parameters, information is also included about yields of quark spectra at $\ptz=40$~GeV, $Y_{q,0}$ (for better readability, in this appendix, we label quantities connected with quark-initiated and gluon-initiated jets by subscripts $q$ and $g$, respectively.). This information is important for the evaluation of jet $\Raa$ using parameterizations with and without nPDFs, which are typically used in the numerator and denominator of $\Raa$, respectively. Since in this case the overall normalization of the jet \pt\ spectra does not divide out in the ratio, one needs to scale the quenched yield in the numerator of $\Raa$ by a factor $(Y_{q,0}^A/Y_{q,0}^B)\cdot (f_{q,0}^B/f_{q,0}^A)$, where superscripts $A$ and $B$ label unquenched jet spectra with and without nPDFs, respectively. Since only ratios of $Y_{q,0}$ need to be used, these factors are provided with no units. 

Table \ref{tab:app:jets0} shows parameterizations of jet \pt\ spectra for inclusive $R=0.4$ jets obtained using PYTHIA8, PYTHIA8 + EPPS16 nPDFs, and HERWIG7. The spectra in $|\eta|<2.8$ with no nPDFs are reweighted before fitting such that they reproduce jet cross-sections measured in $pp$ collisions published in Ref.~\cite{ATLAS:2018gwx}. The spectra with nPDFs are reweighted by the same factors. 
Results for the jet \pt\ spectra without the reweighting are also provided for completeness and labeled ``no-weight''.

Table \ref{tab:app:jets1} shows parameterizations of jet \pt\ spectra for inclusive $R=0.2$ jets and $R=0.2$ $b$-jets obtained using PYTHIA8 and PYTHIA8 + EPPS16 nPDFs. No reweighting to data is applied. 

Table \ref{tab:app:jets2} shows parameterizations of jet \pt\ spectra for jets in $\gamma$-jet events obtained using PYTHIA8, PYTHIA8 + EPPS16 nPDFs, and HERWIG7. The spectra with no nPDFs are reweighted before fitting such that they reproduce jet cross-sections measured in $pp$ collisions published in Ref.~\cite{ATLAS:2023iad}. The spectra with nPDFs are reweighted by the same factors. Two kinds of reweighting are done. In the first case, PYTHIA8 jet spectra from the direct photon sample only are reweighted (labeled in the table as $w_1$). In the second case, PYTHIA8 jet spectra in the sample with fragmentation photons only are reweighted (labeled in the table as $w_2$). In both cases, reweighting by a constant factor is applied first to minimize the $\chi^2$ between the data and MC. 
Results for the jet \pt\ spectra without the reweighting are also provided for completeness and labeled ``no-weight''.

\begin{figure}
\begin{center}
\includegraphics[width=0.45\textwidth]{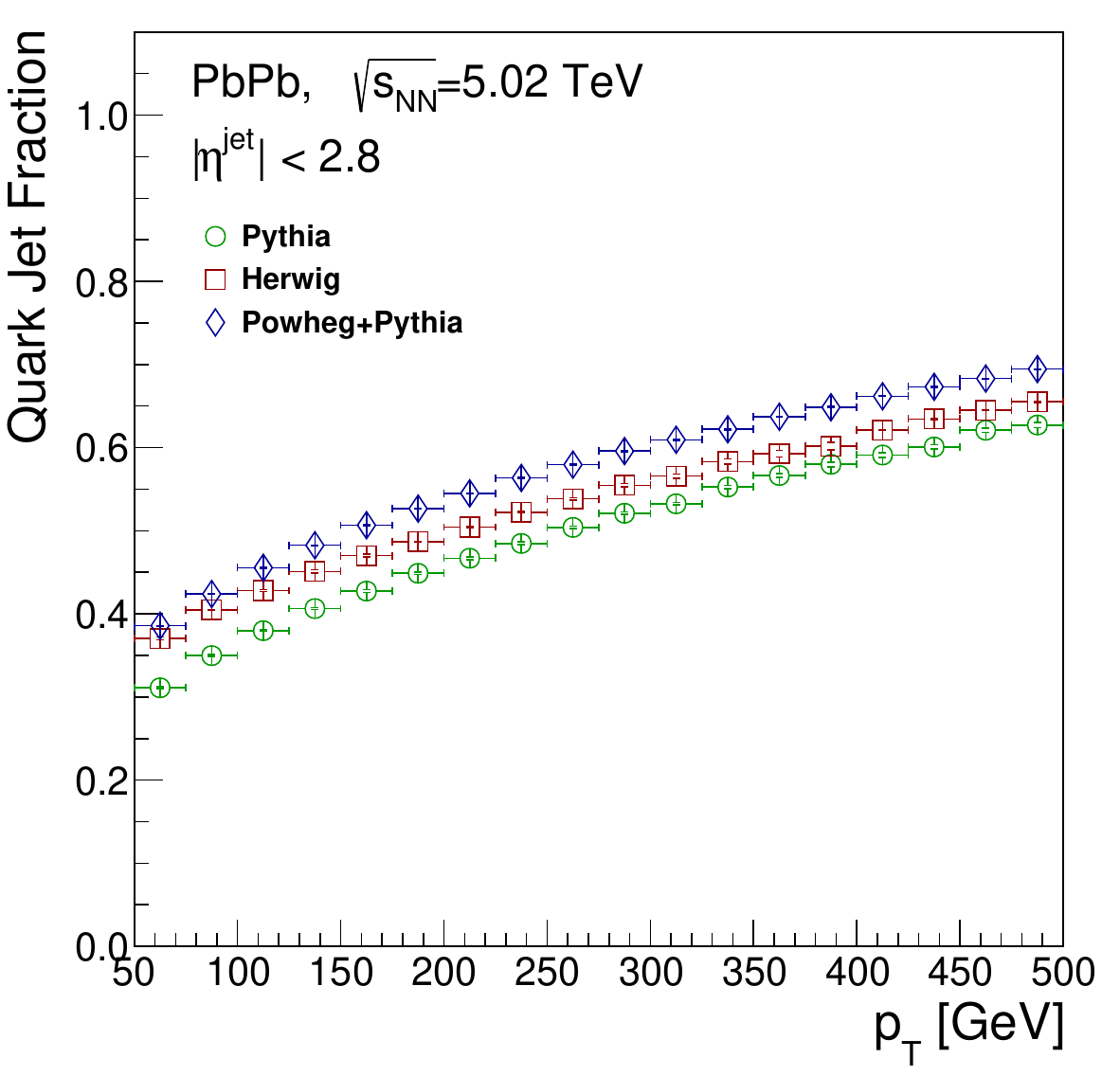}
\includegraphics[width=0.45\textwidth]{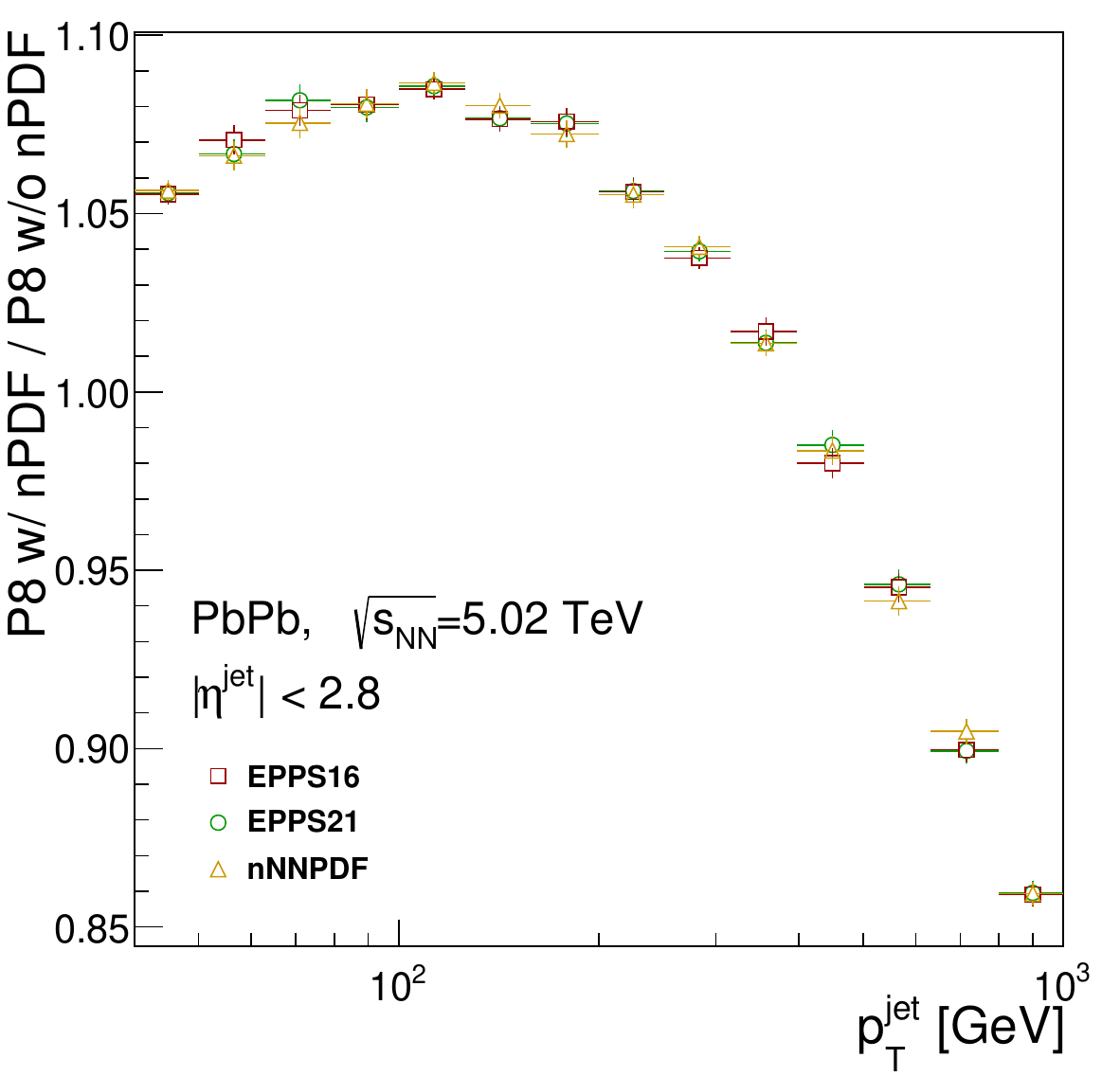}
\end{center}
\caption{\label{fig:appA} \textit{Left:} Comparison of the evolution of quark jet fraction as a function of $\ptjet$ for $R=0.4$ jets obtained using PYTHIA8, HERWIG7, and POWHEG + PYTHIA8. \textit{Right:} Ratio of jet \pt\ spectra obtained using PYTHIA8 with different sets of nPDFs, namely EPPS16, EPPS21, and nNNPDF3.0, to jet spectra from PYTHIA8 with default $pp$ PDFs from tune A14.}
\end{figure}

The important MC ingredient in the detailed description of jet spectra is the quark jet fraction $f_q$ (labeled $f_1$ in the body of the paper). As one can see from the tables, $f_{q,0}$ differs among samples. The left panel of Figure~\ref{fig:appA} shows a comparison of the evolution of quark jet fraction as a function of $\ptjet$ for $R=0.4$ jets obtained using PYTHIA8, HERWIG7, and POWHEG~\cite{Alioli:2010xa} + PYTHIA8. One can see that the $f_q(\ptjet)$ evolution differs between PYTHIA8 and HERWIG7, which is one of the reasons why differences between these two generators are often used to determine some of the uncertainties connected with the jet energy scale \cite{ATLAS:2017bje}. The $f_q(\ptjet)$ in PYTHIA8 and POWHEG+PYTHIA8 differs not by the shape but rather by a shift. In the study presented in this paper, this difference is covered by a much larger variation of the color factor $c_F$ used in jet quenching parameterizations p1-p3 (for details see Section~\ref{sec:incl}). The uncertainty in the modeling of $f_q$ may, however, affect the extracted values of quenching parameters such as $\pars$ or $\hat{q}$. 

Another important ingredient in the detailed jet quenching modeling comes from nPDFs. In this study, we used only one set of nPDFs coming from the EPPS16 parameterization, which was also used within the experimental work in Ref.~\cite{ATLAS:2023iad}. To evaluate the impact of nPDFs on jet \pt\ spectra we evaluated the ratio of jet \pt\ spectra obtained using PYTHIA8 with different sets of nPDFs, namely EPPS16, EPPS21 \cite{Eskola:2021nhw}, and nNNPDF3.0 \cite{AbdulKhalek:2022fyi}, and jet spectra from PYTHIA8 with default $pp$ PDFs from tune A14. The result is plotted in the right panel of Figure~\ref{fig:appA}. One can conclude that within the statistical uncertainties of the simulated sample (25 million events in total for each of the PDF sets), we see an agreement among different sets of nPDFs.

    \begin{table}[h] 
   \footnotesize
   \renewcommand{\arraystretch}{1.8} 
     \centering 
       \begin{tabular}{|c|c|c|c|c|c|c|c|c|c|c|c|c|} 
    \hline 
   
MC & $\eta$ selection & $f_{q,0}$ & $Y_{q,0}$ & $\beta_{0,q}$  &  $\beta_{0,g}$  &  $\beta_{1,q}$  &  $\beta_{1,g}$  &  $\beta_{2,q}$  &  $\beta_{2,g}$  &  $\beta_{3,q}$  &  $\beta_{3,g}$ \\ \hline
 \multirow{1}{*}{\stackanchor{PYTHIA8     }{$R=0.4$ inclusive jets, no-weight}} 
   &  $|\eta|<2.8$ & $0.28$ & \num{0.4} & $4.04$ & $4.54$ & $-1.18$ & $-1.08$ & $-0.51$ & $-0.45$ & $-0.11$ & $-0.11$ \\   \hline 
 \hline 
 \multirow{1}{*}{\stackanchor{PYTHIA8     }{$R=0.4$ inclusive jets}} 
   &  $|\eta|<2.8$ & $0.30$ & \num{0.32} & $4.07$ & $4.49$ & $-1.04$ & $-1.00$ & $-0.41$ & $-0.39$ & $-0.09$ & $-0.10$ \\   \hline 
 \hline 
 \multirow{1}{*}{\stackanchor{PYTHIA8+nPDF}{$R=0.4$ inclusive jets, no-weight}} 
   &  $|\eta|<2.8$ & $0.31$ & \num{0.48} & $4.07$ & $4.53$ & $-1.16$ & $-1.02$ & $-0.49$ & $-0.40$ & $-0.11$ & $-0.10$ \\   \hline 
 \hline 
 \multirow{1}{*}{\stackanchor{PYTHIA8+nPDF}{$R=0.4$ inclusive jets}} 
   &  $|\eta|<2.8$ & $0.29$ & \num{0.31} & $3.99$ & $4.44$ & $-1.08$ & $-1.02$ & $-0.42$ & $-0.37$ & $-0.10$ & $-0.10$ \\   \hline 
 \hline 
 \multirow{1}{*}{\stackanchor{HERWIG7     }{$R=0.4$ inclusive jets}} 
   &  $|\eta|<2.8$ & $0.28$ & \num{0.003} & $3.04$ & $3.92$ & $-2.01$ & $-1.58$ & $-0.76$ & $-0.59$ & $-0.13$ & $-0.12$ \\   \hline 
 \hline 
\end{tabular} 
     \caption{Values of parameterization of jet $\pt$ spectra (see (\ref{eq:unmod}),(\ref{eq:exp})) for  inclusive $R=0.4$ jets.
     Quantities connected with quark-initiated and gluon-initiated jets are labeled by subscripts $q$ and $g$, respectively.
     } 
     \label{tab:app:jets0} 
     \end{table} 
   
   \begin{table} 
   \footnotesize 
   \renewcommand{\arraystretch}{1.8} 
     \centering 
       \begin{tabular}{|c|c|c|c|c|c|c|c|c|c|c|c|c|} 
    \hline 
   
MC & $\eta$ selection & $f_{q,0}$ & $Y_{q,0}$ & $\beta_{0,q}$  &  $\beta_{0,g}$  &  $\beta_{1,q}$  &  $\beta_{1,g}$  &  $\beta_{2,q}$  &  $\beta_{2,g}$  &  $\beta_{3,q}$  &  $\beta_{3,g}$ \\ \hline
 \multirow{1}{*}{\stackanchor{PYTHIA8     }{$R=0.2$ inclusive jets}} 
   &  $|\eta|<2.1$ & $0.31$ & \num{0.17} & $2.30$ & $3.45$ & $-3.63$ & $-2.29$ & $-1.90$ & $-1.11$ & $-0.37$ & $-0.23$ \\   \hline 
 \hline 
 \multirow{1}{*}{\stackanchor{PYTHIA8+nPDF}{$R=0.2$ inclusive jets}} 
   &  $|\eta|<2.1$ & $0.30$ & \num{0.16} & $2.20$ & $3.46$ & $-3.69$ & $-2.23$ & $-1.91$ & $-1.05$ & $-0.37$ & $-0.22$ \\   \hline 
 \hline 
 \multirow{1}{*}{\stackanchor{PYTHIA8     }{$R=0.2$ $b$-jets}} 
   &  $|\eta|<2.8$ & $0.88$ & \num{0.016} & $3.16$ & $2.48$ & $-2.44$ & $-2.35$ & $-1.06$ & $-1.00$ & $-0.21$ & $-0.20$ \\   \hline 
 \hline 
 \multirow{1}{*}{\stackanchor{PYTHIA8+nPDF}{$R=0.2$ $b$-jets}} 
   &  $|\eta|<2.8$ & $0.88$ & \num{0.016} & $3.30$ & $2.68$ & $-2.06$ & $-2.04$ & $-0.78$ & $-0.83$ & $-0.16$ & $-0.18$ \\   \hline 
 \hline 
\end{tabular} 
     \caption{Values of parameterization of jet $\pt$ spectra (see (\ref{eq:unmod}),(\ref{eq:exp})) for  $R=0.2$ inclusive jets and $b$-jets.   
     Quantities connected with quark-initiated and gluon-initiated jets are labeled by subscripts $q$ and $g$, respectively.
     } 
     \label{tab:app:jets1} 
     \end{table} 
   
   \begin{table} 
   \footnotesize 
   \renewcommand{\arraystretch}{1.8} 
      \setlength{\extrarowheight}{3.5pt}
 \centering 
       \begin{tabular}{|c|c|c|c|c|c|c|c|c|c|c|c|c|} 
    \hline 
   
MC & $\eta$ selection & $f_{q,0}$ & $Y_{q,0}$ & $\beta_{0,q}$  &  $\beta_{0,g}$  &  $\beta_{1,q}$  &  $\beta_{1,g}$  &  $\beta_{2,q}$  &  $\beta_{2,g}$  &  $\beta_{3,q}$  &  $\beta_{3,g}$ \\ \hline
 \multirow{1}{*}{\stackanchor{PYTHIA8, $R=0.4$     }{$\gamma$-jets, no-weight}} 
   &  $|\eta|<2.1$ & $0.81$ & \num{0.12} & $-2.76$ & $-1.14$ & $-8.60$ & $-7.21$ & $-4.66$ & $-3.89$ & $-0.96$ & $-0.79$ \\   \hline 
 \hline 
 \multirow{1}{*}{\stackanchor{PYTHIA8, $R=0.4$     }{$\gamma$-jets, $w_1$}} 
   &  $|\eta|<2.1$ & $0.79$ & \num{0.12} & $-3.34$ & $-0.23$ & $-8.70$ & $-3.72$ & $-4.05$ & $-0.60$ & $-0.72$ & $0.06$ \\   \hline 
 \hline 
 \multirow{1}{*}{\stackanchor{PYTHIA8, $R=0.4$     }{$\gamma$-jets, $w_2$}} 
   &  $|\eta|<2.1$ & $0.81$ & \num{0.15} & $-2.41$ & $-1.14$ & $-7.55$ & $-7.21$ & $-3.58$ & $-3.89$ & $-0.65$ & $-0.79$ \\   \hline 
 \hline 
 \multirow{1}{*}{\stackanchor{PYTHIA8+nPDF, $R=0.4$}{$\gamma$-jets, no-weight}} 
   &  $|\eta|<2.1$ & $0.83$ & \num{0.11} & $-2.63$ & $-1.24$ & $-8.34$ & $-7.40$ & $-4.48$ & $-4.02$ & $-0.92$ & $-0.82$ \\   \hline 
 \hline 
 \multirow{1}{*}{\stackanchor{PYTHIA8+nPDF, $R=0.4$}{$\gamma$-jets, $w_1$}} 
   &  $|\eta|<2.1$ & $0.80$ & \num{0.12} & $-3.26$ & $-0.36$ & $-8.60$ & $-3.99$ & $-4.00$ & $-0.81$ & $-0.72$ & $0.00$ \\   \hline 
 \hline 
 \multirow{1}{*}{\stackanchor{PYTHIA8+nPDF, $R=0.4$}{$\gamma$-jets, $w_2$}} 
   &  $|\eta|<2.1$ & $0.82$ & \num{0.14} & $-2.30$ & $-1.24$ & $-7.34$ & $-7.40$ & $-3.43$ & $-4.02$ & $-0.62$ & $-0.82$ \\   \hline 
 \hline 
 \multirow{1}{*}{\stackanchor{HERWIG7,      $R=0.4$}{$\gamma$-jets, no-weight}} 
   &  $|\eta|<2.1$ & $0.87$ & \num{0.022} & $-0.79$ & $3.37$ & $-3.21$ & $-2.28$ & $-0.74$ & $-1.17$ & $-0.10$ & $-0.32$ \\   \hline 
 \hline 
 \multirow{1}{*}{\stackanchor{HERWIG7,      $R=0.4$}{$\gamma$-jets, $w_1$}} 
   &  $|\eta|<2.1$ & $0.80$ & \num{0.23} & $-1.23$ & $2.86$ & $-4.88$ & $-3.34$ & $-1.46$ & $-1.84$ & $-0.12$ & $-0.39$ \\   \hline 
 \hline 
 \multirow{1}{*}{\stackanchor{HERWIG7,      $R=0.4$}{$\gamma$-jets, $w_2$}} 
   &  $|\eta|<2.1$ & $0.78$ & \num{0.21} & $-2.21$ & $3.37$ & $-7.35$ & $-2.28$ & $-3.61$ & $-1.17$ & $-0.71$ & $-0.32$ \\   \hline 
 \hline 
\end{tabular} 
     \caption{Values of parameterization of jet $\pt$ spectra (see (\ref{eq:unmod}),(\ref{eq:exp})) for  $R=0.4$ $\gamma$-jets. 
     Quantities connected with quark-initiated and gluon-initiated jets are labeled by subscripts $q$ and $g$, respectively.
     } 
     \label{tab:app:jets2} 
     \end{table}

\section{Jet quenching parameters}
\label{sec:appB}

In this Appendix, we provide a full list the jet quenching parameters derived in Section~\ref{sec:incl} and discuss the correlation between parameters $\alpha$ and $s$. 

Values of parameters for all jet quenching schemes used in this study are listed in Table~\ref{tab:app:quench}.
Two types of error on parameter $s$ are evaluated. The first one, labeled $\delta s_1$ was obtained by propagating $\ptjet$ uncorrelated systematic uncertainties and statistical uncertainties from jet $\Raa$ measurement in Ref.~\cite{ATLAS:2018gwx} to the minimization. The second one, labeled $\delta s_2$ was obtained by propagating systematic uncertainties fully correlated in $\ptjet$. Due to the normalization used in Equation~(\ref{eq:s}) where $\ptz=40$~GeV, the values of parameter $s$ represent an average energy loss of quark-initiated jet with $\ptjet=40$~GeV in units of GeV.    

Parameters $\alpha$ and $s$ are highly anti-correlated in each centrality interval. The correlation coefficient
\begin{equation}
    \rho \equiv \frac{\mathrm{cov}(\alpha,s)}{\sqrt{\mathrm{var}(\alpha) \mathrm{var}(s)}},
\nonumber
\end{equation}
is smaller than $-0.9$ in centrality bins in the range $0-50\%$. In centrality bins $50-60\%$ and $60-70\%$ $\rho$ is $-0.7$ and $-0.6$, respectively. The degree of anti-correlation can be maximized by replacing $s$ by $\exp(s)$ in parameterization (\ref{eq:s}). Then, $\rho$ is close to $-1$ in all centrality bins, allowing to parameterize the relation between $\alpha$ and $s$ as $\alpha=p_0 + p_1 s$, with $p_0$ and $p_1$ being new parameters, but $p_1=-0.21 \pm 0.01$ universally for all centrality bins. To significantly reduce the correlation between the parameters $\alpha$ and $s$ one can then re-parameterize (\ref{eq:s}) as follows
\begin{equation}
    \pars_i = c_{F,i} ~ \exp(s) ~ (\ptjet)^{\alpha-0.21s}
\nonumber
\end{equation}
where we set $\ptz=1$~GeV for simplicity. With this new parameterization, the $\rho$ is well below $0.5$ for the majority of centrality bins, which may improve the precision in determining the parameters. At the same time, the overall improvement in the description of the jet $\Raa$ data would be rather small given already good values of $\chi^2$ reported in Section~\ref{sec:incl} and the parameterization would introduce centrality dependence in the $\alpha$ parameter and therefore effectively double the number of values needed to describe the data.

   \begin{table} 
   \footnotesize 
   \renewcommand{\arraystretch}{1} 
     \centering 
       \begin{tabular}{|c|c|c|c|c|c|c|c|c|c|c|c|} 
    \hline 

\shortstack{paramete-\\rization} & $\alpha$ & $c_F$ &   &  0-10\%  &  10-20\%  &  20-30\%  &  30-40\%  &  40-50\%  &  50-60\%  &  60-70\%  \\ \hline 
 \multirow{3}{*}{p1} &  \multirow{3}{*}{\stackanchor{$0.27$}{$\pm 0.03$}} & \multirow{3}{*}{$1.78$} &  $s$ &  11.72 &  8.94 &  6.98 &  5.14 &  3.59 &  2.31 &  1.49  \\ \cline{4-11} 
& & & $\delta s_1$ &  0.13 &  0.12 &  0.06 &  0.07 &  0.03 &  0.11 &  0.09  \\ \cline{4-11} 
& & & $\delta s_2$ &  0.15 &  0.21 &  0.26 &  0.37 &  0.45 &  0.58 &  0.74  \\ \hline \hline 
 \multirow{3}{*}{p2} &  \multirow{3}{*}{\stackanchor{$0.24$}{$\pm 0.02$}} & \multirow{3}{*}{$1.31$} &  $s$ &  14.53 &  11.19 &  8.75 &  6.48 &  4.53 &  2.93 &  1.90  \\ \cline{4-11} 
& & & $\delta s_1$ &  0.17 &  0.17 &  0.09 &  0.09 &  0.05 &  0.13 &  0.11  \\ \cline{4-11} 
& & & $\delta s_2$ &  0.19 &  0.26 &  0.32 &  0.47 &  0.56 &  0.73 &  0.94  \\ \hline \hline 
 \multirow{3}{*}{p3} &  \multirow{3}{*}{\stackanchor{$0.29$}{$\pm 0.03$}} & \multirow{3}{*}{$2.25$} &  $s$ &  9.97 &  7.55 &  5.87 &  4.30 &  3.03 &  1.91 &  1.23  \\ \cline{4-11} 
& & & $\delta s_1$ &  0.12 &  0.10 &  0.04 &  0.06 &  0.05 &  0.09 &  0.08  \\ \cline{4-11} 
& & & $\delta s_2$ &  0.13 &  0.19 &  0.22 &  0.32 &  0.38 &  0.51 &  0.66  \\ \hline \hline 
 \multirow{3}{*}{p4} &  \multirow{3}{*}{\stackanchor{$0.33$}{$\pm 0.02$}} & \multirow{3}{*}{$1.78$} &  $s$ &  11.10 &  8.38 &  6.59 &  4.78 &  3.36 &  2.03 &  1.35  \\ \cline{4-11} 
& & & $\delta s_1$ &  0.10 &  0.14 &  0.09 &  0.05 &  0.01 &  0.09 &  0.11  \\ \cline{4-11} 
& & & $\delta s_2$ &  0.19 &  0.26 &  0.32 &  0.47 &  0.56 &  0.73 &  0.93  \\ \hline \hline 
 \multirow{3}{*}{p5} &  \multirow{3}{*}{\stackanchor{$0.30$}{$\pm 0.02$}} & \multirow{3}{*}{$1.78$} &  $s$ &  11.71 &  8.81 &  6.93 &  4.98 &  3.52 &  2.27 &  1.41  \\ \cline{4-11} 
& & & $\delta s_1$ &  0.14 &  0.16 &  0.11 &  0.06 &  0.02 &  0.06 &  0.10  \\ \cline{4-11} 
& & & $\delta s_2$ &  0.15 &  0.21 &  0.26 &  0.37 &  0.48 &  0.53 &  0.72  \\ \hline \hline 
 \multirow{3}{*}{p6} &  \multirow{3}{*}{\stackanchor{$0.40$}{$\pm 0.01$}} & \multirow{3}{*}{$1.78$} &  $s$ &  7.82 &  6.29 &  5.19 &  3.95 &  2.92 &  1.94 &  1.29  \\ \cline{4-11} 
& & & $\delta s_1$ &  0.07 &  0.07 &  0.03 &  0.05 &  0.03 &  0.09 &  0.08  \\ \cline{4-11} 
& & & $\delta s_2$ &  0.22 &  0.31 &  0.37 &  0.53 &  0.70 &  0.77 &  1.04  \\ \hline \hline 
 \multirow{3}{*}{p7} &  \multirow{3}{*}{\stackanchor{$0.34$}{$\pm 0.02$}} & \multirow{3}{*}{$1.78$} &  $s$ &  9.27 &  7.35 &  5.90 &  4.43 &  3.20 &  2.10 &  1.37  \\ \cline{4-11} 
& & & $\delta s_1$ &  0.09 &  0.11 &  0.06 &  0.07 &  0.04 &  0.09 &  0.08  \\ \cline{4-11} 
& & & $\delta s_2$ &  0.20 &  0.27 &  0.34 &  0.48 &  0.59 &  0.75 &  0.96  \\ \hline \hline 
 \multirow{3}{*}{p8} &  \multirow{3}{*}{\stackanchor{$0.15$}{$\pm 0.02$}} & \multirow{3}{*}{$1.78$} &  $s$ &  2.35 &  1.86 &  1.53 &  1.11 &  0.83 &  0.56 &  0.36  \\ \cline{4-11} 
& & & $\delta s_1$ &  0.03 &  0.02 &  0.02 &  0.02 &  0.01 &  0.01 &  0.04  \\ \cline{4-11} 
& & & $\delta s_2$ &  0.10 &  0.14 &  0.17 &  0.24 &  0.29 &  0.37 &  0.48  \\ \hline \hline

 \hline 
\end{tabular} 
     \caption{Jet quenching parameters for parameterizations p1-p8 defined in Section~\ref{sec:mod}. For details see the text. } 
     \label{tab:app:quench} 
     \end{table} 

\section{\npart dependence of jet quenching}
\label{sec:appC}

In this appendix, we discuss the connection between the Glauber path-length and $\npart$ dependence of jet energy loss and the general parameterization of centrality dependence of the energy loss.  

For the Glauber model, we found in Section~\ref{sec:path} that $\pars(\ptjet=100~\mathrm{GeV})= d_0 +d_1 \parl^{\delta}$ with $\delta=2.21 \pm 0.05$, $d_1=0.43\pm0.03$ for $d_0=0$. One can now use values published in Ref.~\cite{Loizides:2017ack} to derive the dependence of $\parl$ on $\npart$ which can be parameterized as
\begin{equation}
   \parl = n_0 + n_1 \npart^\nu.   
\end{equation}
For $n_0$ fixed to zero (to maintain consistency with $d_0=0$) and for centralities $0-80\%$, one obtains for parameters $n_1$ and $\nu$ following values: $n_1=0.60\pm0.01$ and $\nu=0.36\pm0.01$. This implies that within the Glauber model, $\pars$ should depend on $\npart$ as
\begin{equation}
   \pars = m_0 + m_1 \npart^\mu.   
\end{equation}
with $m_0=0$, $m_1=d_1 n_1^\delta$, and $\mu = \nu\delta$. When inserting earlier obtained values, one obtains: $m_1 = 0.14 \pm 0.01$ and $\mu = 0.80 \pm 0.05$. 

\begin{figure}
\begin{center}
\includegraphics[width=0.45\textwidth]{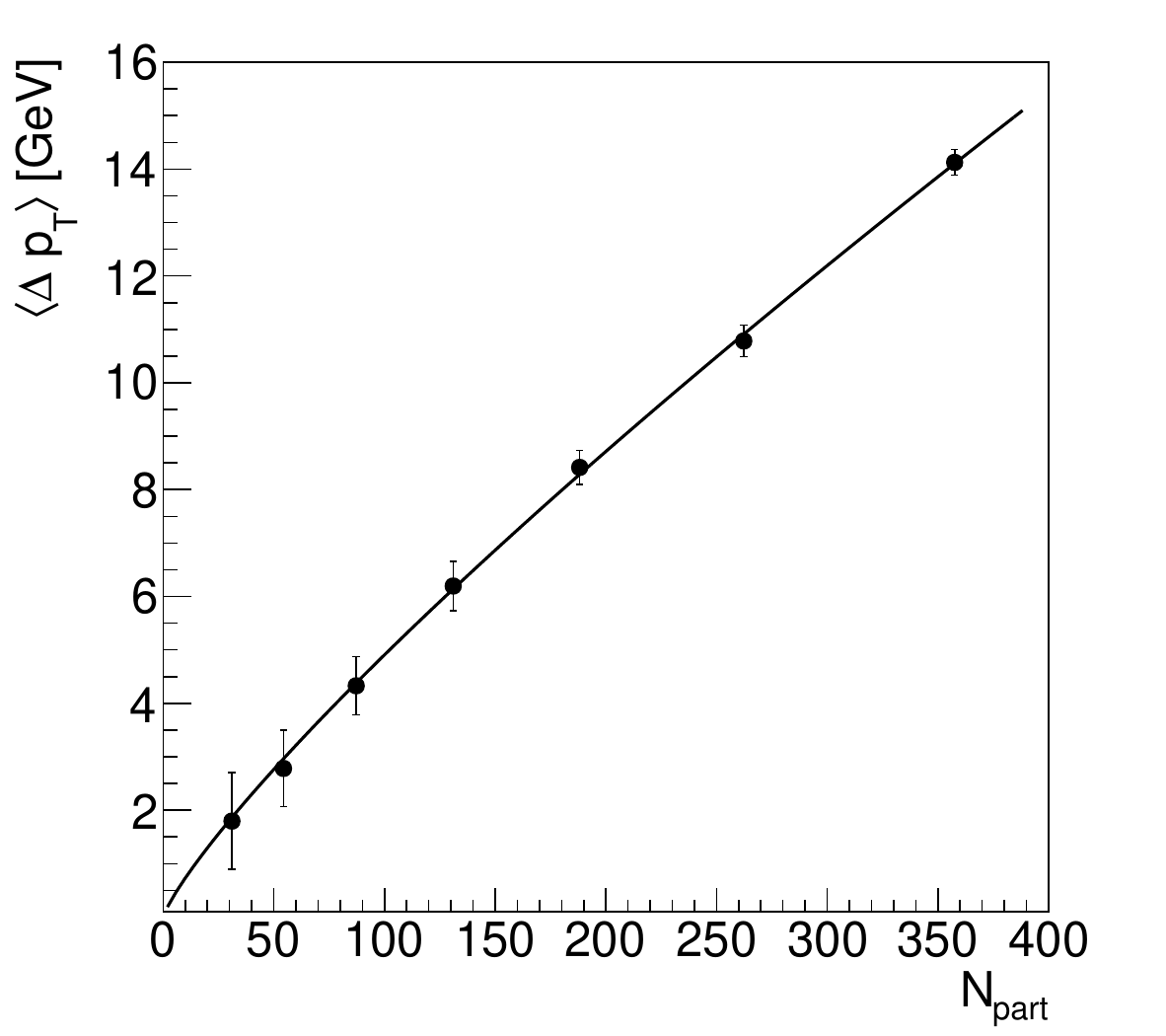}
\end{center}
\caption{\label{fig:appC} 
The $\npart$ dependence of the average energy loss of jets with $\ptjet=100$~GeV. The values of $\pars$ are based on quenching parameterization p1.  
}
\end{figure}

One can compare calculated values of $m_1$ and $\mu$ with a fit of $\pars(\npart)$ dependence obtained directly using values from Tab.~\ref{tab:app:quench} for the default parameterization p1. This fit is shown in Fig.~\ref{fig:appC}. In the fit with $m_0$ unconstrained, the value of $m_0$ was found to be consistent with zero. With $m_0=0$, the other parameters have following values: $m_1=0.11 \pm 0.03$ and $\mu = 0.83 \pm 0.04$. These results, therefore, imply an overall consistency of the parametric modeling.  These results also imply that the average energy loss is not exactly a linear function of $\npart$, as reported in earlier work \cite{Spousta:2016agr}, but it is close to it. 

Obtained parameterizations allow to precisely characterize the measured energy loss in jet $\Raa$ \cite{ATLAS:2018gwx} in all centrality bins using six parameters. Four parameters characterizing the average energy loss: $c_F$, $\alpha$, $\mu$, $m_1'$, and two parameters characterizing its fluctuations, $c_0$ and $c_1$ discussed in Section~\ref{sec:mod}. The connection between the centrality dependent parameter $s$ from (\ref{eq:s}) and parameters $m_1'$ and $\mu$ is following
\begin{equation}
    s = m_1' \npart^\mu,
\end{equation}
where $m_1' = m_1 (40/100)^\alpha$. The factor $(40/100)^\alpha$ comes from the fact that the path-length parameterization was evaluated for the reference value $\ptjet=100$~GeV. While this $\ptjet$ value may be judged as an arbitrary number, we should mention that the results of the path-length parameterization remain the same within uncertainties when changing to other $\ptjet$ values within the $\pt$ range of the jet \Raa measurement \cite{ATLAS:2018gwx}.  

To summarize this appendix, we can provide a suitable parameterization of centrality-dependent quenching of inclusive jets as follows,
\begin{equation}
    \pars = c_F m_1 \npart^\mu \left( \frac{\ptjet}{100~\mathrm{GeV}} \right)^\alpha, 
\end{equation}
with $c_F=1.78$, $\alpha=0.27\pm0.03$, $m_1=0.11 \pm 0.03$, and $\mu = 0.83 \pm 0.04$ which needs to be smeared according to (\ref{eq:fluct3}) with $c_0 = 1.4 \pm 0.2$ and $c_1 = 0.3 \pm 0.1$.

\end{document}